\documentclass[aps,prd,preprintnumbers,amsmath,amssymb,nofootinbib,superscriptaddress,floatfix]{revtex4-2}
\usepackage[latin1]{inputenc}
\usepackage{slashed,xspace,nicefrac}
\usepackage{amsmath,amssymb,amsfonts}
\usepackage{textcomp}
\usepackage{natbib}
\usepackage{enumitem}
\usepackage{array,multirow}
\usepackage{threeparttable}
\usepackage{appendix}
\usepackage[table,xcdraw]{xcolor}
\usepackage{color} 
\usepackage{subcaption}
\usepackage{tikz}
\usepackage{makecell}
\usepackage[normalem]{ulem}
\usepackage[unicode]{hyperref}
\newcommand{\be}{\begin{equation}}
\newcommand{\ee}{\end{equation}}
\newcommand{\bea}{\begin{eqnarray}}
\newcommand{\eea}{\end{eqnarray}}

\newcommand{\hf}{\nicefrac{1}{2}\xspace}
\newcommand{\nn}{\nonumber\\}

\newcommand{\MAD}{\textsc{MadGraph}\xspace}
\newcommand{\madamc}{\textsc{MadGraph5\_aMC@NLO}\xspace}

\newcommand{\Feyncalc}{\textsc{FeynCalc}\xspace}
\newcommand{\Math}{\textsc{Mathematica}\xspace}
\newcommand{\h}{\ensuremath{\mathcal{H}}}
\newcommand{\hb}{\ensuremath{\overline{\mathcal{H}}}}
\newcommand{\tev}{\mbox{TeV}\xspace}
\newcommand{\gev}{\mbox{GeV}\xspace}

\usepackage{graphicx,here}

\DeclareGraphicsExtensions{.pdf,.png,.jpg}
\graphicspath{{./}{figs/}{figures/}}
\begin{document}
\begin{flushleft} 
KCL-PH-TH/2024-{\bf 35}
\end{flushleft}

\title{Impact of resummation on the production and experimental bounds of scalar high-electric-charge objects}

\author{Jean Alexandre}
\email{jean.alexandre@kcl.ac.uk}
\affiliation{Theoretical Particle Physics and Cosmology group, Department of Physics, King's College London, London WC2R 2LS, UK}

\author{Nick E.\ Mavromatos}
\email{nikolaos.mavromatos@cern.ch}
\affiliation{Physics Division, School of Applied Mathematical and Physical Sciences, National Technical University of Athens, 15780 Zografou Campus,
Athens, Greece}
\affiliation{Theoretical Particle Physics and Cosmology group, Department of Physics, King's College London, London WC2R 2LS, UK}

\author{Vasiliki A.\ Mitsou}
\email{vasiliki.mitsou@ific.uv.es}
\affiliation{Instituto de F\'isica Corpuscular (IFIC), CSIC -- Universitat de Val\`encia,
C/ Catedr\'atico Jos\'e Beltr\'an 2, 46980 Paterna (Valencia), Spain}

\author{Emanuela Musumeci}
\email{emanuela.musumeci@ific.uv.es}
\affiliation{Instituto de F\'isica Corpuscular (IFIC), CSIC -- Universitat de Val\`encia,
C/ Catedr\'atico Jos\'e Beltr\'an 2, 46980 Paterna (Valencia), Spain}

\begin{abstract}
 
A one-loop Dyson-Schwinger-like resummation scheme is applied to scalar High-Electric-Charge compact Objects (HECOs), extending previous work on spin-\hf case.  
The electromagnetic interactions of HECOs are considered within the framework of strongly coupled scalar Quantun Electrodynamics.
The resummation amounts to determining non-trivial ultraviolet (UV) fixed points, at which the effective Lagrangian, which will lead to the pertinent predictions on the cross sections, is computed. 
In contrast to the fermionic HECO case, in which the fixed point structure was determined solely by the interactions of the HECOs with the photon field, in the scalar case the 
existence of non-trivial UV fixed points requires the presence of additional strong self interactions among the HECOs. Our resummation scheme, which is notably different from a lattice strong-coupling approach, makes the computation of the pertinent scalar-HECO-production cross sections reliable, thus allowing revisiting the mass bounds obtained from searches for such objects in current or future colliders. Our \MAD implementation of the results leads to enhanced (up to $\sim 30$\%) lower bounds on the mass of scalar HECOs, as compared to those extracted from the tree-level processes typically used in LHC collider searches by ATLAS and MoEDAL experiments. 

\end{abstract}

\maketitle

\section{Introduction}

High Electric Charge Objects (HECOs), of various spins, are actively searched upon at the LHC~\cite{ATLAS:2013cab,CMS:2013czn,ATLAS:2015hau,ATLAS:2015tyu,ATLAS:2018imb,ATLAS:2019wkg,MoEDAL:2021mpi,ATLAS:2023zxo,ATLAS:2023esy,MoEDAL:2023ost}, while prospects for future colliders have been studied thoroughly~\cite{Altakach:2022hgn}. Such objects arise in a plethora of theoretical models beyond the standard model, ranging from non-strange quark matter aggregates~\cite{Holdom:2017gdc}, 
strangelets~\cite{Farhi:1984qu}, 
Q-balls~\cite{Coleman:1985ki,Kusenko:1997si}, black-hole remnants in models of large extra dimensions~\cite{Koch:2007um}, Schwinger dyons~\cite{Schwinger:1969ib}, electrically charged scalars in neutrino-mass models~\cite{Hirsch:2021wge}, and doubly charged massive particles~\cite{MoEDAL:2014ttp}. Lacking a concrete fundamental theoretical framework of these avatars of new physics, the interpretation of the current searches is based on the use of generic tree-level Drell-Yan (DY) and Photon-Fusion (PF) production processes. However, in view of the non-perturbative nature of the pertinent couplings of the HECOs (large electric charge), truncation to such lowest-order graphs in perturbation theory is unreliable.   

In a previous work~\cite{Alexandre:2023qjo,Musumeci:2024erp}, we have proposed a resummation scheme for quantum corrections, based on the Dyson-Schwinger equations (DS), in strongly coupled quantum electrodynamics. The resummation leads to self-consistent relations between the dressed quantities, the latter depending on a transmutation scale $k$ introduced by dimensional regularization. The 
resummation allows for a non-trivial ultraviolet (UV) fixed point. 
The corresponding dressed mass is proportional to $k$ and therefore formally goes to infinity, 
but the dressed coupling constants defining this fixed point are finite.
This model has been proposed in \cite{Alexandre:2023qjo} to describe a heavy HECO particle of spin-\hf, with a non-perturbative electric charge.\footnote{In a different setting, a similar in spirit one-loop improved resummation has also been proposed~\cite{Alexandre:2019iub} to describe the gauge effective field theory dynamics of magnetic monopoles, which also constitute non-perturbative field-theoretic objects, due to the large value of the magnetic charge, as a result of the Dirac quantization condition~\cite{Baines:2018ltl}. Such resummations should be compared to lattice gauge-theory approaches, which started becoming available for monopoles~\cite{Farakos:2024ggp}. However, for the case of magnetic monopoles, if the latter have structure, as is the case of the magnetic monopoles that characterize grand-unification extension of the standard model~\cite{Mavromatos:2020gwk}, there is significant suppression of the associated production cross section at collider, as a result of the underlying structure~\cite{Drukier:1981fq}. Thus, our resummation scheme are rather more appropriate for point-like Dirac monopoles~\cite{Alexandre:2019iub}.} Specifically, at the non-trivial UV fixed point, the effective Lagrangian describing the electromagnetic interactions of the fermionic-HECO $\psi$ with the photon field $A_\mu$, is given by (in the Feynman gauge):
\bea\label{efflagF}
{\cal L}_\text{eff}\to\frac{1}{2}A_\mu\left(\eta^{\mu\nu}\Box -\frac{\omega^\star}{1+\omega^\star}\partial^\mu\partial^\nu\right)A_\nu +\overline\psi\left( i\slashed\partial+\frac{g}{\sqrt{1+\omega^\star}}\slashed A-\frac{M}{\mathcal Z^\star}\right)\psi~,
\eea
where $\omega$ and $\mathcal Z$ denote the wavefunction renormalization  of the photon and fermionic HECO, respectively, while the asterisk ($\star$) indicates values at the fixed point. $M$ is the (renormalized) mass of the HECO, while $g$ is its electric charge. The fixed point quantities $\omega^\star, \mathcal Z^\star$ are all found to be of order one. As stressed in \cite{Alexandre:2023qjo},
naively, from the effective Lagrangian \eqref{efflagF}, at the UV fixed point, 
one would expect to use the tree-level vertex rule $\Gamma_\mu = \frac{g}{\sqrt{1 + \omega^\star}} \, \gamma_\mu$ for the HECO--photon vertex. 
 However, the Lagrangian \eqref{efflagF} is \emph{gauge fixed} due to the non-trivial $\omega^\star$-dependent longitudinal terms ($\propto \partial^\mu \partial^\nu$) of the gauge sector. Hence, the standard Ward identity stemming from gauge invariance in conventional Quantum Electrodynamics (QED), which would imply the equality of the wave-function renormalizations for the HECOs ($\mathcal Z$) and the vertex ($\mathcal Z_V$), $\mathcal Z = \mathcal Z_V$, no longer applies in the resummed strongly coupled QED~\cite{Alexandre:2023qjo}. Thus, one can define a \emph{renormalized} HECO coupling $g_R$ (at the UV fixed point) by the rescaling: 
$g \to g_R\, \sqrt{(1 + \omega^\star)}\, \frac{{\mathcal Z}^\star}{{\mathcal Z_V}^\star}$,  
which leads to 
$g \to \frac{1}{\mathcal Z_V^\star} g_R$. We expect on natural grounds that  ${\mathcal Z^\star_V}^{-1} \sim \mathcal Z^\star \sim 1$, which leads to the 
following Feynman rules for the resummed QED, including a generalized non-trivial 
generalized vertex rule (omitting, from now on, the suffix $R$ for notational brevity):
\bea\label{feynmanrules}
G^\text{eff}&=&i\frac{\slashed p+\mathcal M(\Lambda)}{p^2-\mathcal M(\Lambda)^2}\nonumber\\
\Delta^\text{eff}_{\mu\nu}&=&\frac{-i}{q^2}\left(\eta_{\mu\nu}+ \omega^\star\, \frac{q_\mu q_\nu}{q^2}\right)\nn
\Gamma^\text{eff}_\mu&=&g \, \mathcal Z^\star\gamma_\mu~,
\eea
respectively (in the Feynman gauge), where $\mathcal M \equiv M/\mathcal Z^\star$, and the UV fixed point has been defined as the limit where the transmutation mass scale $k \to \Lambda$, an UV cutoff scale that defines the range of validity of the effective theory \eqref{efflagF}. The (renormalized) HECO coupling $g$ can be expressed in terms of the (physical) electron charge as $g=n \,e$, where $n \in \mathbb Z$.
We stress that, in this resummation scheme, to avoid overcounting of the resummed loops, we need to restrict ourselves to tree-level DY or PF production processes of the type discussed in Refs.~\cite{Song:2021vpo,MoEDAL:2021mpi,ATLAS:2023esy}, 
but with the important replacement of the standard tree-level Feynman rules by \eqref{feynmanrules}. 

Such improved DS-like resummation schemes allow for a consistent derivation of cross sections for the production of such objects, especially HECOs, at colliders, which in turn implies the extraction of reliable mass bounds of these avatars of new physics beyond the Standard Model (SM) from experimental data.
In the current work, we extend the approach of \cite{Alexandre:2023qjo} to the case of scalar (spin-0) HECOs. Surprisingly, although in the fermion-HECO case the existence of a resummation UV fixed-point, at large transmutation scales, $k \to \Lambda$, is secured by considering solely the electromagnetic interactions of the HECO with the photon, however, as we shall demonstrate, in the scalar-HECO case this requires sufficiently strong self interactions among the HECOs, whose self coupling has to be significantly larger than the square of the gauge coupling. Once the existence of an UV fixed point is established, the effective renormalized mass of the scalar HECO is derived, by solving the appropriate resummation equations, and this can be used to fit the appropriate experimental data, thus extracting reliable mass bounds on HECOs from current collider searches. 

The structure of the article is the following: in the next section \ref{sec:2} we describe the DS-like improved one-loop resummation scheme
for the case of spin-0 massive HECOs with non-trivial quartic self interactions. 
In Section~\ref{sec:3}, we discuss the implementation of the resummed effective model of scalar HECOs into realistic \MAD-UFO models. 
In Section~\ref{sec:4}, we compute the relevant cross sections for the production of the scalar HECOs at colliders, within our resummation scheme, which in Section~\ref{sec:5} lead to the extraction of reliable  bounds for the scalar HECO masses, by comparing them with the current  upper bounds on production cross sections set by ATLAS~\cite{ATLAS:2019wkg,ATLAS:2023esy} and MoEDAL~\cite{MoEDAL:2021mpi,MoEDAL:2023ost} experiments. Finally, conclusions and outlook are presented in Section~\ref{sec:6}. Lastly, some technical details of our approach, including the calculation of loop integrals using dimensional regularization, and derivation of the expressions for the gauge-boson polarization tensor, scalar self energy and scalar self interaction, are given in Appendices. 

\section{Non-perturbative resummation approach to scalar HECOs }\label{sec:2}

We embark now on our one-loop resummation for spin-$0$ HECOs. 
First we remark that lacking a concrete microscopic model embedding the scalar HECOs into the SM framework, one should resort to an effective field theory description. 
For instance, the usual assumption that the scalar HECO has only U(1) hypercharge interactions with the SM sector~\cite{Altakach:2022hgn}, leads to trilinear derivative couplings of HECO pairs with the $Z$-boson and photon.
In such models one has contributions of the $Z$-boson, along with the photon, to DY production process. In ref.~\cite{Song:2021vpo}, however, which the experimental searches interpretation were based upon~\cite{ATLAS:2019wkg,ATLAS:2023esy,MoEDAL:2021mpi,MoEDAL:2023ost}, such $Z$-boson contributions to DY have been ignored. 

\begin{figure}[ht]
\centering
\begin{subfigure}[b]{0.25\textwidth}
         \centering
         \includegraphics[width=\textwidth]{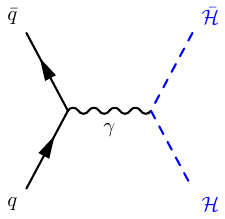}
         \caption{Drell-Yan, $\gamma$ exchange}
         \label{fig:dy-gamma}
     \end{subfigure}
     \hspace{0.04\textwidth}
     \begin{subfigure}[b]{0.25\textwidth}
         \centering
         \includegraphics[width=\textwidth]{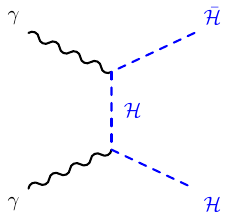}
         \caption{Photon fusion, $t$ channel}
         \label{fig:pf-tchan}
     \end{subfigure}
     \hspace{0.04\textwidth}
     \begin{subfigure}[b]{0.25\textwidth}
         \centering
         \includegraphics[width=\textwidth]{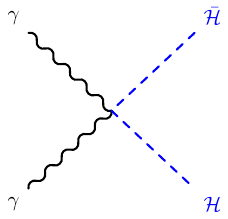}
         \caption{Photon fusion, four-point}
         \label{fig:pf-four}
     \end{subfigure}
    \caption{Considered processes for the production of scalar HECO-anti-HECO (\h\hb) pairs at hadron colliders: (a) Drell-Yan with only $\gamma$ exchange; (b) $t$-channel photon fusion; and (c) four-point photon fusion. We do not indicate arrows in the dashed lines for charged HECOs for brevity. The $u$-channel photon-fusion diagram is also considered but not shown explicitly.}
    \label{fig:dy-pf-diagrams}
\end{figure}

The leading-order diagrams on which the resummation is applied are presented in Fig.~\ref{fig:dy-pf-diagrams}. For the purposes of studying the effects of resummation on the production of scalar HECOs via DY and PF processes, 
and comparing directly our predictions to the results of the pertinent collider searches, we shall consider in this work only the coupling of scalar HECOs to photons, via a scalar-electrodynamics effective Lagrangian. Hence, the $Z$-boson is not included in the DY process, as shown in Fig.~\ref{fig:dy-gamma}. We also ignore $Z$-boson fusion in the PF process, as shown in Figs.~\ref{fig:pf-tchan} and~\ref{fig:pf-four}, which is suppressed at the LHC~\cite{Altakach:2022hgn,Alexandre:2023qjo}. Potential contributions from the $Z$-boson will not affect qualitatively our main conclusion, namely that our resummation leads to significantly enhanced HECO mass bounds (see also~\cite{Alexandre:2023qjo}).  As we shall see, to ensure the existence of non-trivial UV stable fixed points, which lies at the core of our resummation, it will be necessary to include scalar-HECO self interactions.  We develop the relevant approach in the next three  subsections, \ref{sec:sqed}, \ref{sec:ressqed} and \ref{sec:fpsqed}, where we discuss the model of strongly coupled scalar electrodynamics with scalar self interactions, its one-loop Dyson-Schwinger resummation, and the resulting UV fixed-point structure, and dressed HECO masses, respectively.

\subsection{Review on Scalar Quantum Electrodynamics}\label{sec:sqed}

The starting point is the bare Lagrangian
\be\label{action}
{\cal L}_\text{bare}=\frac{1}{2}A^\mu\left[\eta_{\mu\nu}\Box-(1-\lambda)\partial_\mu\partial_\nu\right]A^\nu
+D_\mu\phi(D^\mu\phi)^\ast-m^2\phi\phi^\ast-\frac{h}{4}(\phi\phi^\ast)^2~,
\ee
where $D_\mu=\partial_\mu+igA_\mu$ and $\lambda$ is a gauge fixing parameter. We assume the following form for the dressed Lagrangian
\cite{Itzykson:1980rh}
\be\label{Ldressed}
{\cal L}_\text{eff}=\frac{1}{2}A^\mu\left[\eta_{\mu\nu}\Box-(1-\lambda)\partial_\mu\partial_\nu\right]A^\nu
+\frac{\omega}{2}A^\mu\left[\eta_{\mu\nu}\Box-\partial_\mu\partial_\nu\right]A^\nu
+{\mathcal Z}^2 D_\mu\phi(D^\mu\phi)^\ast-M^2\phi\phi^\ast-\frac{H}{4}(\phi\phi^\ast)^2~,
\ee
where only the transverse part of the gauge field gets quantum corrections, and the dressed scalar kinetic term respects gauge invariance. 
The dressed scalar and gauge boson propagators are then given by the following expressions
\bea\label{props}
G(p)&=&\frac{i}{{\mathcal Z}^2p^2-M^2}~,\\
D_{\mu\nu}(q)&=&\frac{-i}{(1+\omega)q^2}\left(\eta_{\mu\nu}+\frac{1+\omega-\lambda}{\lambda}\frac{q_\mu q_\nu}{q^2}\right)~,\nonumber
\eea
respectively, 
and the dressed gauge vertices are 
\bea\label{vertex}
\mbox{3-point}&~&~~~~ig{\mathcal Z}(p_\mu^1\pm p_\mu^2)~,\\
\mbox{4-point}&~&~~~~2ig^2{\mathcal Z}^2\eta_{\mu\nu}~.\nonumber
\eea
Rescaling the fields as $A_\mu\to A_\mu/\sqrt{1+\omega}$ and $\phi\to\phi/{\mathcal Z}$ leads then to the canonically normalized Lagrangian
\be\label{efflag}
\tilde{\cal L}_\text{eff}=\frac{1}{2}A^\mu\left[\eta_{\mu\nu}\Box-\frac{1-\lambda+\omega}{1+\omega}\partial_\mu\partial_\nu\right]A^\nu
+\left(\partial_\mu\phi+i\tilde gA_\mu\phi\right)\left(\partial^\mu\phi^\ast-i\tilde gA^\mu\phi^\ast\right)
-\tilde M^2\phi\phi^\ast-\frac{\tilde H}{4}(\phi\phi^\ast)^2~,
\ee
where the effective parameters are
\be\label{gmeff}
\tilde g\equiv\frac{g}{\sqrt{1+\omega}}~~,~~\tilde H\equiv\frac{H}{{\mathcal Z}^4}~~,~~\tilde M^2\equiv\frac{M^2}{{\mathcal Z}^2}~.
\ee
The above rescaling assumes that the Ward identities are satisfied, and we will question this assumption further down,
where the effective theory is treated like a bare theory in a preferred gauge (see Sec.\ref{sec:3}).

\subsection{One-loop resummation of (strongly coupled) scalar electrodynamics}\label{sec:ressqed}

We assume that the quantum corrections $M^2-m^2, {\mathcal Z}-1, \omega, H$ are momentum independent. 
These corrections will depend on the scale $k$ introduced by dimensional regularization though, 
and we will confirm the consistency of the scale independence of the corrections by proving the existence of a fixed point in the flow equations with $k$.

The resummation scheme we base our study on is the following. 
We consider one-loop-like (resummed) Feynman graphs for quantum corrections to scalar and gauge propagators, as well as the scalar self-coupling,
but we replace the bare propagators and vertices by the dressed ones. This is a bit similar to the DS non-perturbative approach, 
but does not involve two-loop-like graphs. Because of the non-perturbative feature of this approach, 
the coupling $h$ and $g^2$ do not need to be small, which is the motivation of this work.
The resulting set of self-consistent equations is derived in $d=4-\epsilon$ dimensions (see details in Appendices~\ref{app:loop}--\ref{app:selfinter}) and is 
\bea\label{DSeqs}
1-\frac{m^2}{M^2}&=&\frac{1}{8\pi^2{\mathcal Z}^2}\left(\frac{g^2}{\lambda}+\frac{H}{{\mathcal Z}^2}\right)\frac{1}{\epsilon}\left(\frac{k\mathcal Z}{M}\right)^\epsilon~,\\
{\mathcal Z}^2&=&1+\frac{g^2}{8\pi^2}\frac{1}{\epsilon}\left(\frac{k\mathcal Z}{M}\right)^\epsilon\left(\frac{3}{1+\omega}-\frac{1}{\lambda}\right)~,\nn
\omega&=&\frac{g^2}{24\pi^2{\mathcal Z}^2}\frac{1}{\epsilon}\left(\frac{k\mathcal Z}{M}\right)^\epsilon~,\nn
H&=&h+\frac{3}{8\pi^2}\left(-\frac{H^2}{{\mathcal Z}^4}+\frac{g^4}{\lambda^2}(2{\mathcal Z}^2-1)
+\frac{Hg^2}{{\mathcal Z}^2\lambda}\right)\frac{1}{\epsilon}\left(\frac{k\mathcal Z}{M}\right)^\epsilon~,\nonumber
\eea
where finite terms are omitted and the scale $k$ is a running transmutation mass scale arising  from dimensional regularization. Keeping in mind that $g$ and $h$ do not depend on $k$, we obtain 
\bea
\partial_k\left({\mathcal Z}^2\left(\frac{g^2}{\lambda}+\frac{H}{{\mathcal Z}^2}\right)^{-1}\left(1-\frac{m^2}{M^2}\right)\right)&=&
\frac{1}{8\pi^2}\left(\frac{k\mathcal Z}{M}\right)^{-1}\partial_k\left(\frac{k\mathcal Z}{M}\right)~,\\
\partial_k\left(\frac{\lambda(1+\omega)}{3\lambda-1-\omega}({\mathcal Z}^2-1)\right)
&=&\frac{g^2}{8\pi^2}\left(\frac{k\mathcal Z}{M}\right)^{-1}\partial_k\left(\frac{k\mathcal Z}{M}\right)~,\nn
\partial_k({\mathcal Z}^2\omega)&=&\frac{g^2}{24\pi^2}\left(\frac{k\mathcal Z}{M}\right)^{-1}\partial_k\left(\frac{k\mathcal Z}{M}\right)~,\nn
\partial_k\left(\left(-\frac{H^2}{{\mathcal Z}^4}+\frac{g^4}{\lambda^2}(2{\mathcal Z}^2-1)+\frac{Hg^2}{{\mathcal Z}^2\lambda}\right)^{-1}(H-h)\right)
&=&\frac{3}{8\pi^2}\left(\frac{k\mathcal Z}{M}\right)^{-1}\partial_k\left(\frac{k\mathcal Z}{M}\right)~.\nonumber
\eea
Taking into account the following boundary conditions at $k=m$
\be
M(m)=m~,~~~~\mathcal Z(m)=1~,~~~~\omega(m)=0~,~~~~H(m)=h~,
\ee
we finally obtain the set of self-consistent equations satisfied by the dressed quantities $M,{\mathcal Z},\omega,H$, which all depend on the scale $k$,
\bea\label{RGflow}
1-\frac{m^2}{M^2}&=&\frac{g^2}{8\pi^2{\mathcal Z}^2}\left(\frac{1}{\lambda}+\frac{H}{g^2{\mathcal Z}^2}\right)\ln\left(\frac{k\mathcal Z}{M}\right)~, \\
{\mathcal Z}^2&=&1+\frac{g^2}{8\pi^2}\frac{3\lambda-1-\omega}{\lambda(1+\omega)}\ln\left(\frac{k\mathcal Z}{M}\right)~, \nn
\omega&=&\frac{g^2}{24\pi^2{\mathcal Z}^2}\ln\left(\frac{k\mathcal Z}{M}\right)~, \nn
H&=&h+\frac{3}{8\pi^2}\left(-\frac{H^2}{{\mathcal Z}^4}+\frac{g^4}{\lambda^2}(2{\mathcal Z}^2-1)
+\frac{Hg^2}{{\mathcal Z}^2\lambda}\right) \ln\left(\frac{k\mathcal Z}{M}\right)~.\nonumber
\eea
We next proceed to discuss the emergence of an UV fixed point, in the (formal) limit $k \to \infty$.

\subsection{Fixed point and dressed HECO mass}\label{sec:fpsqed}

We are looking for a fixed-point $(\omega^\star,{\mathcal Z}^\star,H^\star)$ solution of Eqs.~\eqref{RGflow} in the limit $k\to\infty$, 
with fixed $k/M$ such that the logarithms remain finite. 
Such a fixed point, if it exists, will be consistent with the assumption that the quantum corrections
we consider are momentum-independent.

Rearranging Eqs.~\eqref{RGflow}, we can eliminate the logarithms and take the limit $M\to\infty$ (for finite $m$) to obtain 
\bea\label{kM}
({\mathcal Z}^\star)^2&=&\frac{\lambda(1+\omega^\star)}{\lambda+(3-8\lambda)\omega^\star+3(\omega^\star)^2}~,\\
1&=&3\omega^\star\left(\frac{1}{\lambda}+\frac{H^\star}{g^2({\mathcal Z}^\star)^2}\right)~,\nn
H^\star&=&h+9\omega^\star\left(-\frac{(H^\star)^2}{g^2({\mathcal Z}^\star)^2}
+\frac{g^2}{\lambda^2}({\mathcal Z}^\star)^2(2({\mathcal Z}^\star)^2-1)+\frac{H^\star}{\lambda}\right)~.\nonumber
\eea
If we assume a solution of these equations for $g^2\ll h$, as well as a finite gauge parameter $\lambda$, we find the approximate solution
\be\label{rangeom}
\omega^\star\simeq\frac{4g^2}{3h}~~~,~~~
({\mathcal Z}^{\star})^2\simeq1+\frac{4g^2}{h}(3-1/\lambda)~~~,~~~
H^\star\simeq\frac{h}{4}~,
\ee
where terms neglected are of higher order in $g^2/h$ and also depend on $\lambda$. 
Hence the gauge independence of $\omega^\star$ and $H^\star$ is satisfied at first order in $g^2/h\ll1$, 
and the gauge dependence of ${\mathcal Z}^\star$ is weak (a mild gauge dependence is to be expected in truncated DS systems, as the current one).  
We finally note that the correction $H^\star-h\simeq-3h/4$ is not necessarily perturbative, which is allowed in the present framework.

Some comments are now in order. First of all, we note that the requirement of real $\mathcal Z$, that is $(\mathcal Z^\star)^2 > 0$, implies (from \eqref{rangeom}) constraints on the allowed values of the gauge parameter. In addition,   
unitarity of the $S$ matrix requires ${\mathcal Z}^\star\ge1$ \cite{Itzykson:1980rh}. If we consider the limiting case $\lambda \to \infty$, then the unitarity requirement would imply 
\begin{align}\label{g2hpositive}
g^2/h > 0\,.
\end{align}
On account of gauge invariance of physically relevant quantities, then, such a condition should characterises all gauge parameters, that is it would be valid for all allowed $\lambda$, that we should obtain the condition 
\begin{align}\label{gaugecond}
\lambda > \frac{1}{3}\,, 
\end{align} 
thus excluding the gauge ($\lambda \to 0$) in our truncation. 

Positivity of $(\mathcal Z)^2$ in a generic gauge \eqref{gaugecond}, with $\omega^\star >0$ (as follows from \eqref{g2hpositive}), requires the denominator in the expression of the first equation \eqref{rangeom} to be positive, that is:
\begin{align}
\mbox{either}~~~~ 0\le\omega^\star<\Big(8\lambda-3 -\sqrt{64\lambda^2 - 60\lambda+9}\Big)/6 
~~~~\mbox{or}~~~~\omega^\star>\Big(8\lambda-3 +\sqrt{64\lambda^2 - 60\lambda+9}\Big)/6~. 
\end{align}
The unitarity requirement $\mathcal Z^\star > 1$ would formally amount to $0 < \omega^\star < 3\lambda - 1$.

From the third of Eqs.~\eqref{RGflow}, we then obtain the effective (physical)  mass of the scalar HECO 
in the limit \eqref{kuv} as:
\bea\label{mass2}
\tilde M =  \Lambda\, \exp\Big(-\frac{32\pi^2}{h}\Big[1 + \frac{4\, g^2}{h}\Big(3-\frac{1}{\lambda}\Big)\Big]\Big)\,,
\eea
We shall absorb any fictitious gauge dependence (which is the unavoidable  consequence of the nature of our resummation scheme, due to a partial graph selection) in the definition of $\Lambda$ so that the physical mass is gauge invariant, and is given by:
\bea\label{mass3}
\tilde M =  \Lambda\, \exp\Big(-\frac{32\pi^2}{h}\Big)~.
\eea

This approximate gauge invariance allows us to pick up a convenient gauge, where we shall perform our analysis of the pertinent HECO-production cross sections. In choosing a convenient gauge we are motivated by the so-called Pinch Technique (PT)~\cite{pinch1,pinch2,pinch3,pinch4}.
Applied initially to 
non-Abelian gauge theories,  the PT shows that gauge-independence of scattering amplitudes is equivalent to
selecting a particular subclass of graphs from the theory using the Feynman gauge $\lambda=1$. This result comes about due to a subtle connection of PT with the background field method of quantizing gauge field theories, which is a special gauge-fixing procedure. The PT essentially assures, that, to all orders in perturbation theory, the gauge-fixing parameter-independent effective n-point correlation functions of the theory, constructed by means of the PT (starting from any gauge-fixing scheme), coincide with the corresponding background n-point functions when the latter are computed at a special value of the corresponding gauge parameter, which is the background Feynman gauge $\lambda=1$. 

 Given that in our case we are dealing with electromagnetic (Abelian) interactions, we mention that 
the validity of PT approach has also been demonstrated for Abelian gauge theories in \cite{pinch5}, in the context of scalar electrodynamics (see also the Appendix of \cite{Mavromatos:2010ar} for a PT study in a version of QED). Hence, in what follows, taking advantage of the aforementioned approximate gauge independence of our fixed-point structure \eqref{rangeom}, 
we shall fix the gauge to the Feynman one $\lambda=1$, which is the physical gauge according to the PT framework. 

For $\lambda=1$, 
positivity of 
$({\mathcal Z}^\star)^2$ would imply that 
a consistent fixed point solution requires:
\be
\mbox{either}~~~~ 0\le\omega^\star<(5-\sqrt{13})/6\simeq0.23~~~~\mbox{or}~~~~\omega^\star>(5+\sqrt{13})/6\simeq1.43~. 
\ee
In addition, the $S$ matrix-unitarity 
condition ${\mathcal Z}^\star\ge1$ \cite{Itzykson:1980rh}, would imply $\omega^\star\le2$. 
The existence of a consistent fixed point therefore requires us to focus on a regime where
\be\label{omegaconditions}
\mbox{either}~~~~0\le\omega^\star\lesssim0.23~~~~\mbox{or}~~~~1.43\lesssim\omega^\star\le2~.
\ee
One can then express $g^2/h$ and $H^\star/h$ as functions of $\omega^\star$ and we find
\bea
g^2/h&=&\frac{3\omega^\star(1-5\omega^\star+3(\omega^\star)^2)^2}
{(1+\omega^\star)(4-50\omega^\star+189(\omega^\star)^2-549(\omega^\star)^3+243(\omega^\star)^4)}~,\\
H^\star/h&=&\frac{(1-3\omega^\star)(1-5\omega^\star+3(\omega^\star))}
{4-50\omega^\star+189(\omega^\star)^2-549(\omega^\star)^3+243(\omega^\star)^4}~.\nonumber
\eea
We can see numerically that the only consistent range for $\omega$, satisfying the conditions (\ref{omegaconditions}) and 
such that $g^2/h>0$, as well as $H^\star>0$, is
\be\label{omega}
0\le \omega^\star\lesssim 0.11~.
\ee
For $\lambda=1$, \eqref{rangeom} becomes:
\be\label{rangeom2}
\omega^\star\simeq\frac{4g^2}{3h}~~,~~~~
({\mathcal Z}^{\star})^2\simeq1+\frac{8g^2}{h}~~,~~~~
H^\star\simeq \frac{h}{4}~,
\ee
which, we stress again, is valid only in the situation where $g^2\ll h$. On account of \eqref{omega}, \eqref{rangeom2} we have in this case:
\bea\label{gvsh}
\frac{g^2}{h} \lesssim 0.0825\, \quad \Rightarrow \quad  h \gtrsim 12.12\,g^2\,. 
\eea
These inequalities should be viewed as gauge independent. 
Thus, for consistency with the existence of a fixed point solution, one must have sufficiently strong self interactions among the scalar HECOs in this model, which needs to be contrasted with the case of spin-\hf HECO~\cite{Alexandre:2023qjo}, where fermion-HECO self interactions were not necessary.

We now regularize the UV limit $k \to \infty$ by 
replacing infinity with the (large but finite) UV cutoff of the effective theory, $\Lambda>0$, which represents the energy scale below which the effective field theory is valid, that is we define the UV fixed point as the limit of the running scale:
\bea\label{kuv}
k \to k_{\rm UV}  = \Lambda\,.
\eea
The effective (physical) gauge-invariant mass of the scalar HECO 
in the limit \eqref{kuv} is given by \eqref{mass3}. In our analysis below we assume the 
saturation~\cite{Musumeci:2024erp} of the constraint \eqref{gvsh} between the interaction couplings. Hence this implies for the mass
\bea\label{gvsh2}
\tilde M \simeq \Lambda\, \exp\Big(-\frac{2.64\,\pi^2}{g^2}\Big)\,,
\eea
which will be used in the analysis of the experimental searches below.

\section{MADGRAPH Implementation - effective Feynman rules}\label{sec:3}

Given the form of the photon propagator (\ref{props}) for $\lambda=1$, 
the fixed point described above is similar to a bare theory with the effective gauge choice 
\be
\lambda_\text{eff}=\frac{1}{1+\omega^\star}~,
\ee
as a result of the partial resummation provided by our non-perturbative approach. 
This prescription corresponds to a ``preferred gauge'', which plays a role in physical predictions, and it effectively prevents us from deriving Ward identities. 
The rescaling $\phi\to\phi/Z$ of the HECO field therefore does not necessarily cancel the charge correction $g\to gZ^\star$, as it would do in standard QED. The reasons are the same as in the fermion HECO case~\cite{Alexandre:2023qjo}, discussed above \eqref{feynmanrules}. In what follows, therefore, we keep this vertex correction for the HECO charge, which we define via the identity
\be
\frac{g_\text{HECO}}{\sqrt{1+\omega^\star}}\equiv(\tilde g {\mathcal Z})^\star~~~~\to~~~~g_\text{HECO}=g{\mathcal Z}^\star~.
\ee
As in the fermionic HECO case~\cite{Alexandre:2023qjo}, we can express the scalar HECO electromagnetic coupling 
$g$ in terms of the physical electron charge as 
\begin{align}\label{ge}
g = n \, e, \quad n \in \mathbb Z ~,
\end{align}
with  $|n| \gtrsim 11$~\cite{Alexandre:2023qjo}, as appropriate for resummation.

The Feynman rules describing the UV fixed point \eqref{rangeom2},  
with the condition 
\eqref{gvsh}
are then:
\bea\label{feynrul}
{\rm Scalar-photon-vertex:} & \qquad &-ig{\mathcal Z}^\star \simeq -ig \sqrt{1 + \frac{8g^2}{h}}~, \\
{\rm Scalar-self-interaction~vertex:} &\qquad & -i\frac{H^\star}{{\mathcal Z}^{\star 4}} 
\simeq -i\frac{h}{4}\left(1 + \frac{8g^2}{h}\right)^{-2}~,\nonumber \\
{\rm Gauge~boson~propagator:} &\qquad & \frac{-i}{p^2 - i\epsilon} \Big(\eta_{\mu\nu} + \omega^\star~ \frac{p_\mu p_\nu}{p^2}\Big) = \frac{-i}{p^2 - i\epsilon} \left(\eta_{\mu\nu} + \frac{4g^2}{3h}\, \frac{p_\mu p_\nu}{p^2}\right)\,,   \quad  \epsilon \to 0^+ ~, \nn
{\rm Charged~scalar~propagator:} &\quad &  \frac{i}{p^2 - \tilde M^2 + i\epsilon}~, 
\quad  \epsilon \to 0^+ ~, \nonumber 
\eea 
where the scalar HECO mass $\tilde M$ is given by \eqref{gvsh2}, which, as already mentioned, follows
from the concrete 
assumption~\cite{Musumeci:2024erp} of the {\it saturation} of the inequality of the constraint \eqref{gvsh}, 
\begin{align}\label{satgvsh}
g^2 \simeq 0.0825\, h \,,
\end{align}
and also make the simplifying assumption, 
following~\cite{Song:2021vpo}, that, on account of gauge invariance in the SM sector, in DY processes of HECO pair production, the scalar HECOs, unlike the spin-\hf HECO, interact only with the photon and not the $Z^0$ boson of the Standard Model. On the other hand, 
the fusion of two $Z^0$ bosons, can produce a pair of HECOs via the quadralinear interactions. However, such processes are suppressed at LHC, as discussed in \cite{Alexandre:2023qjo}, hence in what follows we shall only consider the HECO electromagnetic interactions, assuming that the scalar HECO carries zero hypercharge.

The implementation of the aforementioned resummation effects has been specifically developed for Universal Feynrules Output (UFO) models~\cite{ufo,ufo2}. These models are designed to be compatible with Monte Carlo event generators, such as \madamc~3.5.4~\cite{madgraph}, which is the tool used in this study. The main objective is to simulate the production mechanisms of DY and PF involving spin-0 HECOs at the LHC, utilizing the resummation scheme described in the previous sections.
The developed model considers the interaction with the photons $\gamma$, making it suitable for simulating both DY and PF processes. It employs the Feynman rules~\ref{feynrul}, derived from the effective Lagrangian~\ref{gmeff}, at the UV fixed point~\ref{rangeom}, under the assumption \eqref{satgvsh}.
This UFO model is intended to serve as a tool for comparing experimental results with theoretical benchmarks. A brief description  of the validation procedure and results are provided in Appendix~\ref{validation_appendix}. The UFO model has been validated by comparing simulation-estimated cross-section values against the outcomes of analytical calculations performed using \Math~13.0.1. 

As for the spin-\hf case, the UFO model describing the resummation effects for spin-0 HECOs incorporates two new parameters: (i) The charge multiplicity, denoted by \( n \), which appears in the coupling definition \( g = ne \), with \( e \) being the electron charge, and (ii) the cutoff energy scale parameter, denoted by \( \Lambda \), as defined in Eq.~\eqref{kuv}. The charge and the cutoff parameter directly influence the HECO mass and, thus, affect the calculated cross-section value. Additionally, for the particular center-of-mass energies examined in this study ---namely those at the LHC--- the fine-structure constant is adjusted to \(\alpha_\text{EM} = 1/127.94\), corresponding to the value relevant for the \(Z^0\)-boson mass scale. It is emphasized that this choice notably impacts the cross-section value.

\section{Cross sections for scalar HECO production at Colliders}\label{sec:4}
\begin{figure}[!ht]  
    \centering
    \includegraphics[width=0.495\textwidth]{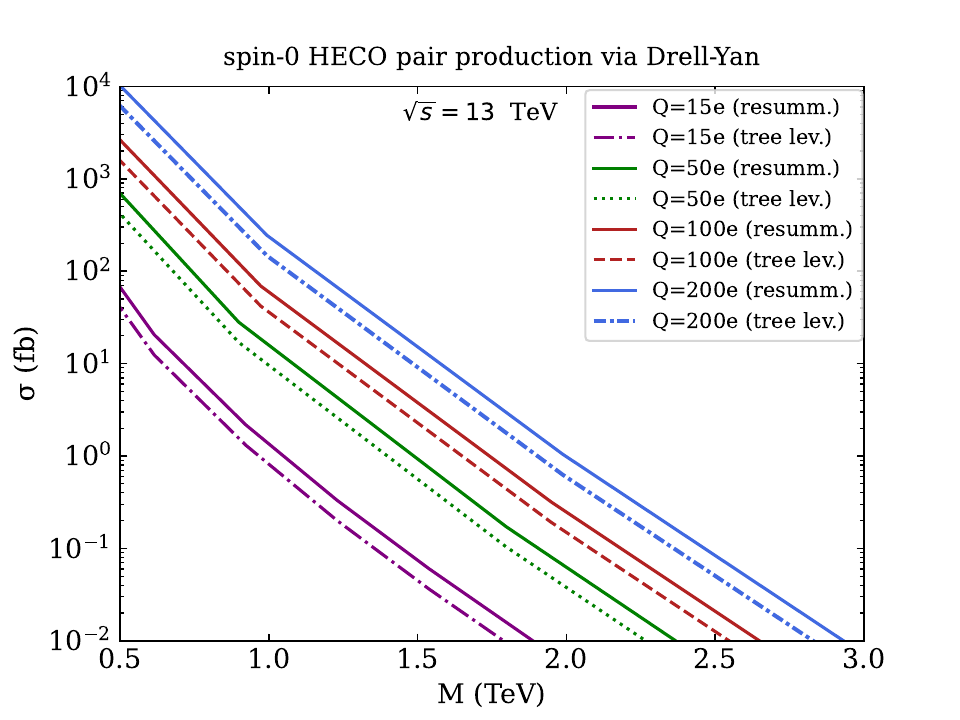} 
    \includegraphics[width=0.495\textwidth]{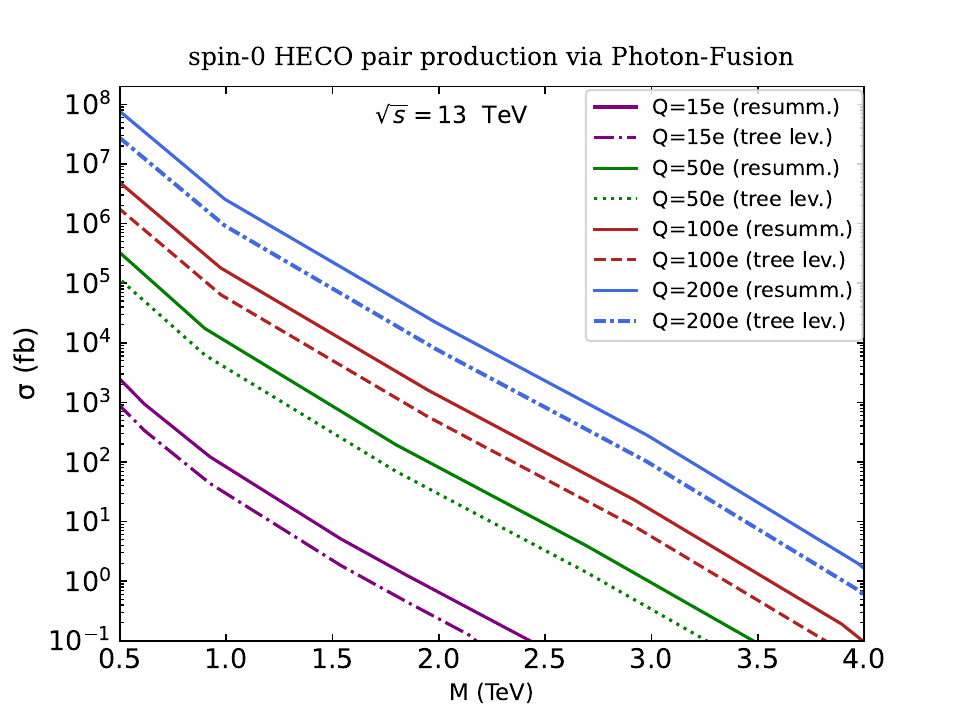}
    \caption{Comparison of cross section values obtained after (solid line) and before (dashed-line) resummation as a function of the scalar HECO mass $M$ for electric charges $Q=15e$, $50e$, $100e$ and $200e$.} 
    \label{fig:aftervsbefore}
\end{figure}

A comparison of the cross-section values obtained at the tree level and after incorporating the resummation effects are reported in Table~\ref{tabDYtreelevsresum} for DY and Table~\ref{tabPFtreelevsresum} for PF. The computations are carried out considering proton--proton ($pp$) collisions with a center-of-mass energy of $\sqrt{s}=13~\tev$. For the DY process the calculations utilize the \texttt{NNPDF23}~\cite{NPDF} parton distribution function (PDF), while for the PF process the \texttt{LUXqed17} PDF~\cite{luxqed} is employed.
When comparing the cross-section values before and after incorporating resummation effects, it is observed that for the same mass ---bare at the tree level and ${\cal M}(\Lambda)$ for the resummation--- the cross section remains within the same order of magnitude. Regarding the impact of DS resummation, an increase by a factor of $\sim$ 1.66 (corresponding to ${{\mathcal Z}^\star}^2$) in the cross-section value for the DY process is observed, consistent with cross-section being proportional to the square of the coupling ${g}^2$, which is $(g{\mathcal Z}^\star)^2$ for the resummation case. On the other hand, for the PF process we have a stronger increase by $\sim 2.76$ (corresponding to ${{\mathcal Z}^\star}^4$) in the cross section as it is proportional to ${g}^4$ . This behavior is also evident in Fig.~\ref{fig:aftervsbefore} where the cross section versus the HECO mass $M$ is drawn before and after resummation at $\sqrt{s}=13~\tev$ for various charge values.  
Hence, in the spin-0 case ---where contributions from the 4-point vertex are also present--- we observe an enhancement in the cross section. This behavior agrees with the spin-\hf case analyzed in~\cite{Alexandre:2023qjo}.  Furthermore, as expected, the cross sections for HECO pair production in the scalar case are lower than those in the spin-\hf one.

\begin{table}[h!]
\caption{Cross-section comparison for scalar HECO production in $pp$ collisions at $\sqrt{s}=13~\tev$ via the DY process between tree level ($\sigma_\text{tree-level}$) and after resummation ($\sigma_\text{resum}$) with $\Lambda= 2~\tev$. The HECO mass is also listed in the last column. The \texttt{NNPDF23} PDF~\cite{NPDF} is used.}
\label{tabDYtreelevsresum}
\centering
\begin{tabular}{ >{\centering}p{2.5cm}>{\centering}p{2.5cm}>{\centering}p{2.5cm}>{\centering\arraybackslash}p{2.5cm} >{\centering\arraybackslash}p{2.5cm} }
\hline\hline
\multicolumn{5}{ c }{DY $p p \to  \h\hb$ @ $\sqrt{s}=13~\tev$, $\Lambda= 2~\tev$} \\
\hline
$Q~(e)$ & $\sigma_\text{tree-level}$ (fb) & $\sigma_\text{resum}$ (fb) & $\sigma_\text{resum}/\sigma_\text{tree}$  & $M$ (\tev) \\
\hline
\phantom{0}20 & 1.173  & 1.943 & 1.66  &  1.030 \\
\phantom{0}60 & 0.110 & 0.182 & 1.65 & 1.857 \\
100 & 0.192 & 0.320 & 1.67  & 1.948 \\
140 & 0.332 & 0.551  & 1.66 & 1.973 \\
180 & 0.519  & 0.863 & 1.66 &  1.984 \\
220 & 0.756 & 1.254  & 1.66 & 1.989\\
\hline\hline
\end{tabular} 
\end{table}

\begin{table}[h!]
\caption{Cross-section comparison for scalar HECO production in $pp$ collisions at $\sqrt{s}=13~\tev$ via the photon-fusion process between tree level ($\sigma_\text{tree-level}$) and after resummation ($\sigma_\text{resum}$) with $\Lambda= 2~\tev$. The HECO mass is also listed in the last column.  \texttt{LUXqed17} PDF~\cite{luxqed} is employed for this process.}
\label{tabPFtreelevsresum}
\centering
\begin{tabular}{ >{\centering}p{2.5cm}>{\centering}p{2.5cm}>{\centering}p{2.5cm}>{\centering\arraybackslash}p{2.5cm} >{\centering\arraybackslash}p{2.5cm}  }
\hline\hline
\multicolumn{5}{ c }{PF $ \gamma \gamma \to \h\hb$ @ $\sqrt{s}=13~\tev$, $\Lambda= 2~\tev$}  \\
\hline
$Q~(e)$ & $\sigma_\text{tree-level}$ (fb) & $\sigma_\text{resum}$ (fb) &  $\sigma_\text{resum}/\sigma_\text{tree}$ & $M$ (\tev) \\
\hline
\phantom{0}20 & $0.748 \times 10^2$  & $2.056 \times 10^2$  & 2.75 & 1.030 \\
\phantom{0}60 & $0.111 \times 10^3$  & $0.306 \times 10^3$ & 2.76 & 1.857\\
100 & $0.576 \times 10^3$ &  $1.591 \times 10^3$ & 2.76& 1.948\\
140 & $1.987 \times 10^3$ & $5.476 \times 10^3$& 2.76 & 1.973 \\
180 & $5.18\phantom{0} \times 10^3$  & $14.30 \times 10^3$ & 2.76&  1.984  \\
220 & $11.31 \times 10^3$ & $31.16 \times 10^3$ & 2.76& 1.989 \\
\hline\hline
\end{tabular} 
\end{table}

\begin{figure}[ht]
\includegraphics[width=.5\textwidth ]{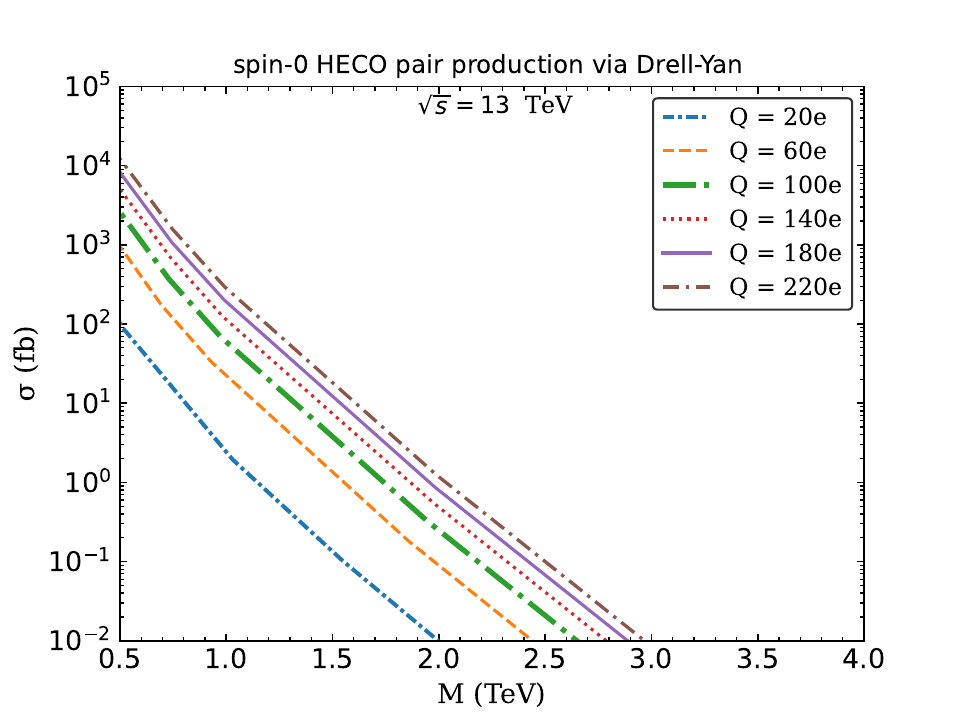}\hfill
\includegraphics[width=.5\textwidth]{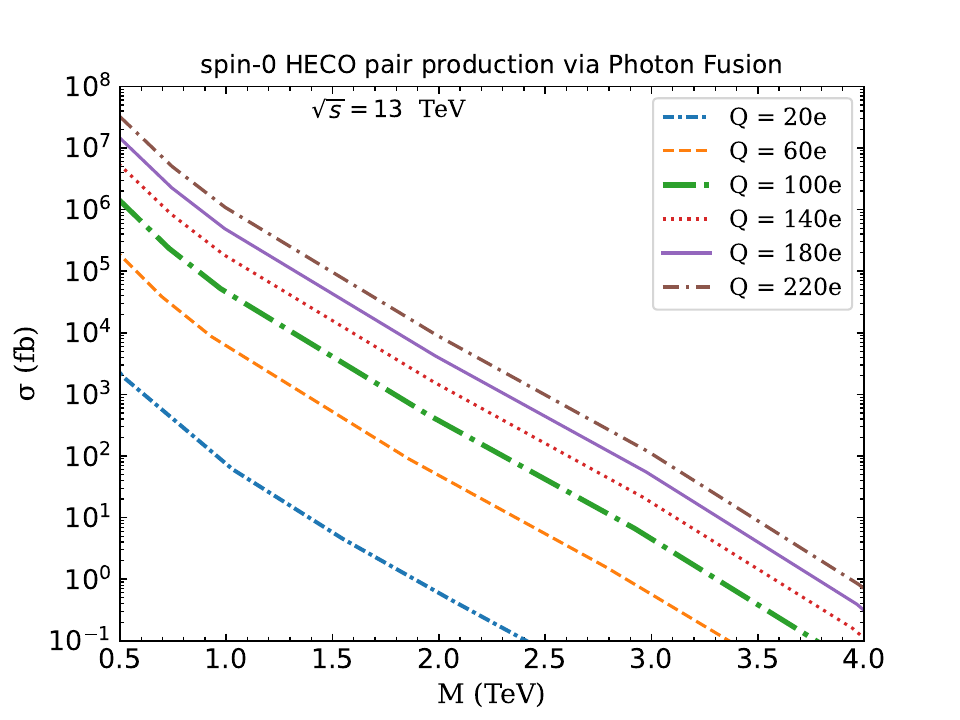}
\\[\smallskipamount]
\includegraphics[width=.5\textwidth]{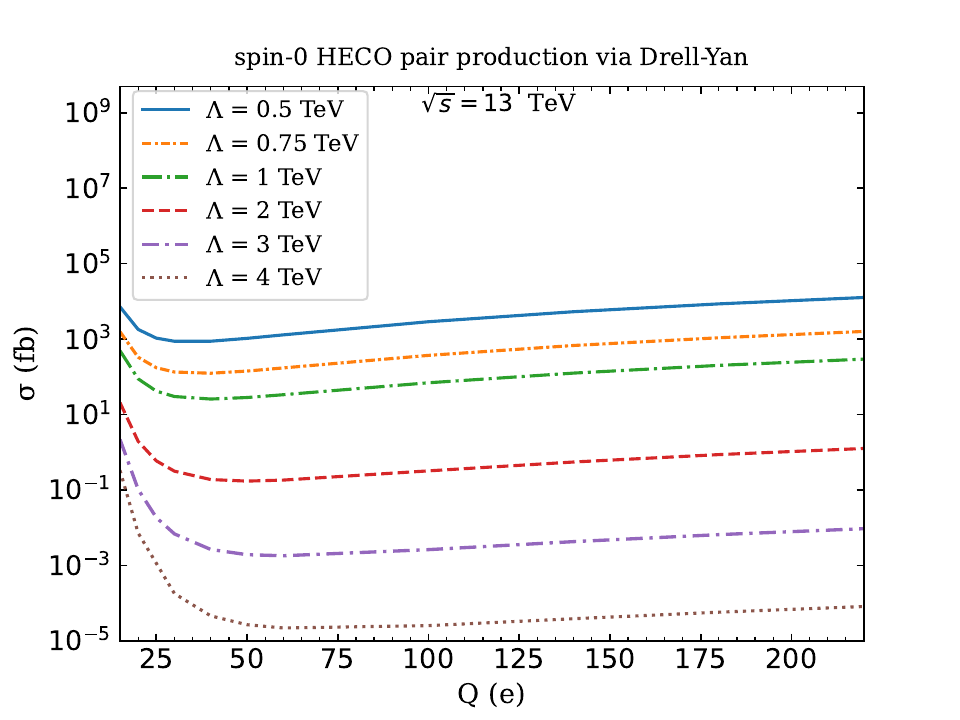}\hfill 
\includegraphics[width=.5\textwidth]{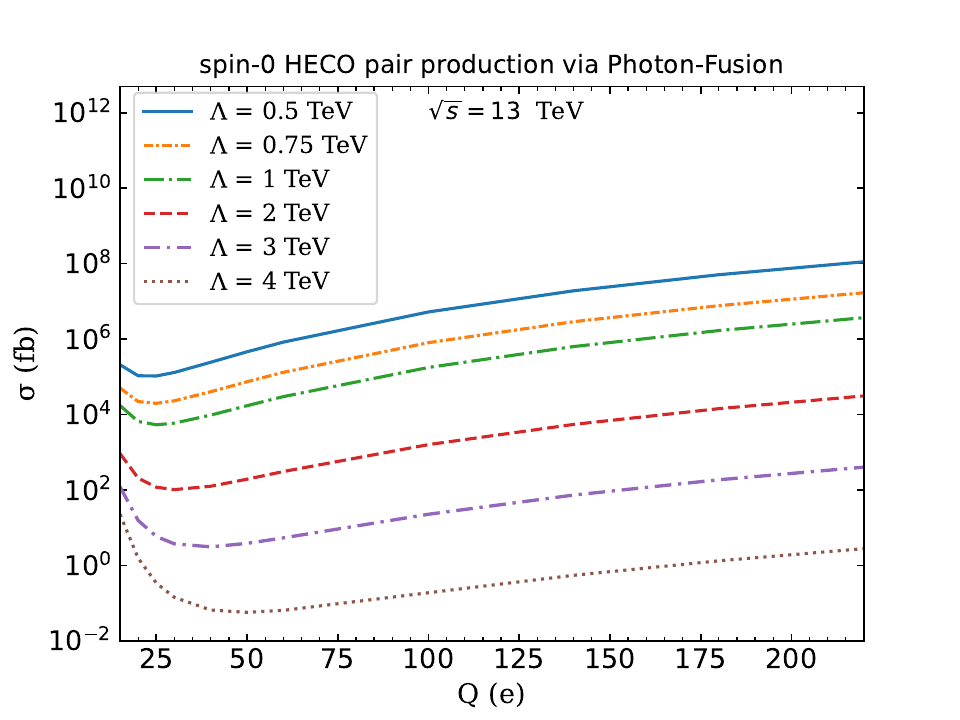}
\includegraphics[width=.5\textwidth]{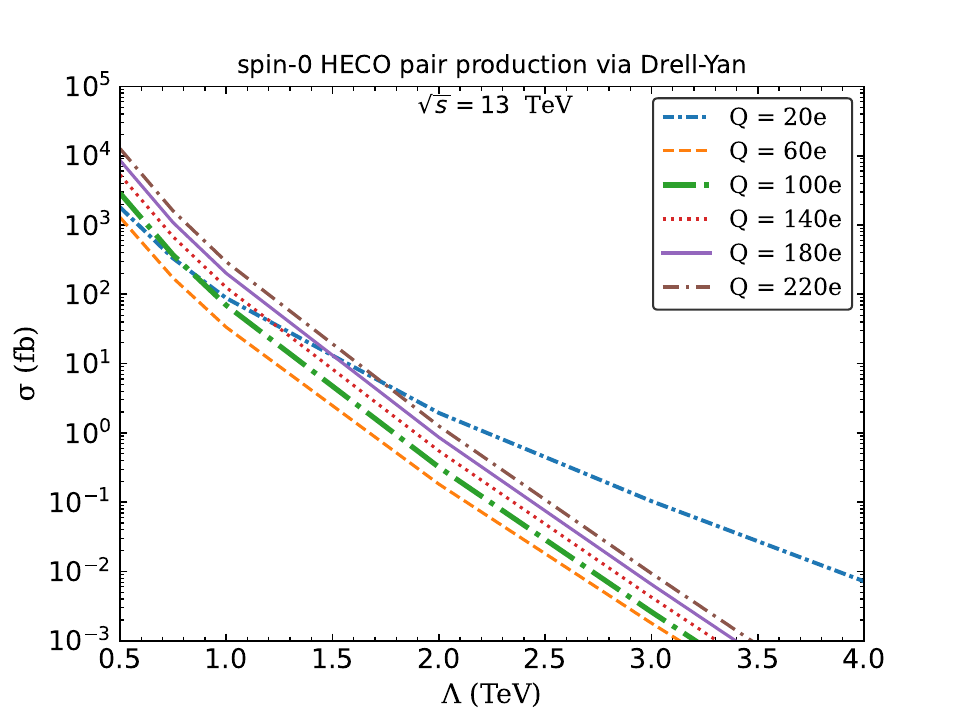}\hfill  
\includegraphics[width=.5\textwidth]{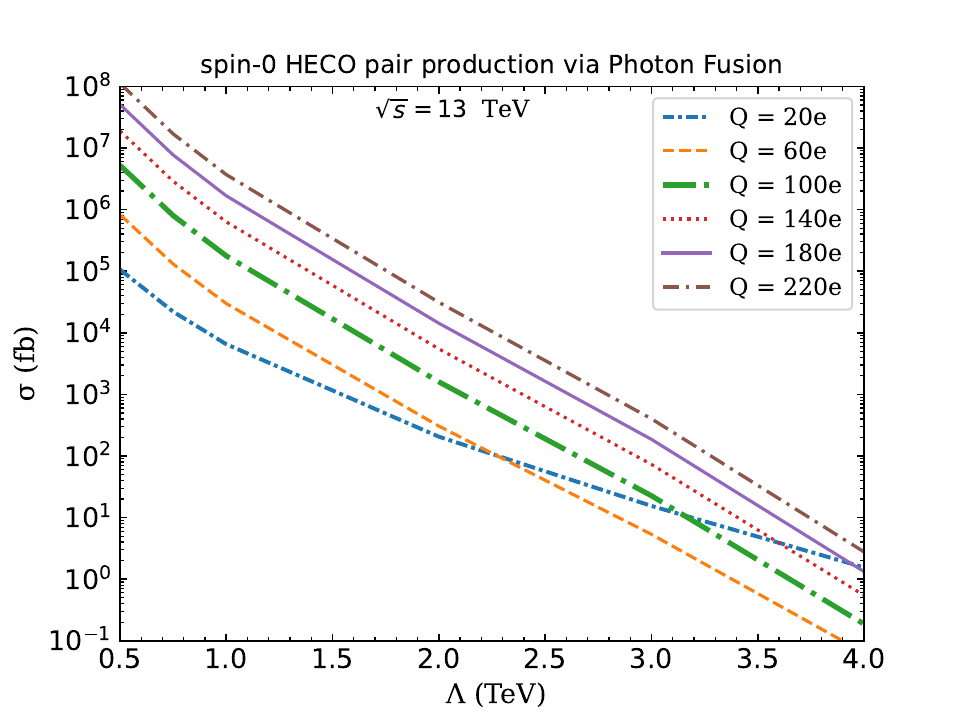}
\\[\smallskipamount]
    \caption{Cross-section values for DY (left) and PF (right) after resummation. The scalar HECO pair-production cross section  at $\sqrt{s}=13~\tev$ is drawn as a function of: (top) the HECO mass $M$ for various charge values; (center) the charge $Q$ for different cutoff $\Lambda$ values and (bottom) the cutoff $\Lambda$ for various $Q$ values. For DY (PF) the \texttt{NNPDF23} (\texttt{LUXqed17})  PDF is utilized.}
    \label{fig:DY_PF_plots}
\end{figure}

Figure~\ref{fig:DY_PF_plots} illustrates the cross-section values for DY (left column) and PF (right column) production mechanisms as a function of various parameters. Similarly to the spin-\hf case, the cross section increases as $Q$ becomes larger for a given HECO mass $M$. However, the behavior varies with different charges; the cross section decreases rapidly as $Q$ increases until it reaches a charge value determined by $\Lambda$. After this point, the cross section experiences a slight increase. This pattern illustrates the nonlinear relationship between $\Lambda$ and charge as analytically described in Eq.~\eqref{gvsh2}.
As $n$ (or $Q$) increases, the HECO mass term contributes significantly, causing a reduction in the production rate. This is why the cross section for $Q = 20e$ is larger than for higher charges when the cutoff scale surpasses a critical value of $\Lambda$. Additionally, it is important to highlight that PF consistently shows higher production levels than the DY process at the collision energy of 13~\tev.

\section{Re-evaluated scalar-HECO mass Limits}\label{sec:5}

In this section, we set new reliable mass limits for scalar HECOs from searches conducted by the ATLAS~\cite{ATLAS:2008xda} and MoEDAL~\cite{MoEDAL:2014ttp} collaborations at the LHC, incorporating the production cross-section predictions obtained in the previous section, which account for resummation effects. We take into account results at $\sqrt{s}=13$ \tev~\cite{ATLAS:2019wkg, MoEDAL:2023ost} in which both DY and PF are examined. We also consider previous searches at $\sqrt{s}=8$ \tev ~\cite{ATLAS:2015tyu,MoEDAL:2021mpi} in which DY was studied. 

Improved and \emph{reliable} scalar-HECO mass limits acquired by DS resummation are presented in the columns ``RES'' of Table~\ref{tab:limits-scalar} for latest ATLAS and MoEDAL analyses and both DY and PF production processes using the cross sections calculated in Section~\ref{sec:4}. We observe that the updated limits are consistently more stringent than the originally obtained by the experimental collaborations, in agreement with the higher cross sections obtained when resummation effects are added compared to the tree-level production rates, as demonstrated in the previous section. The increase in the mass limits spans a wide range of values up to 30\%, as the cross-section upper limits show an irregular pattern depending on the HECO mass and the experiment. For completeness, we also present updated limits for spin-\hf HECOs in Table~\ref{tab:limits-fermion}, including constraints derived from the latest MoEDAL results~\cite{MoEDAL:2023ost}, which provide tree-level limits for HECO pair production.

\begin{table*}[ht]
\caption{95\% CL experimental mass limits for spin-0 HECOs with (RES) and without (LO) resummation techniques to calculate production cross sections. Reported bounds are not necessarily extracted from the same dataset size.}
\label{tab:limits-scalar}
\centering
\begin{tabular}{ c r c c c c c c }
\hline\hline
\multicolumn{8}{ c }{Experimental lower limits at 95\% CL on spin-0 HECO mass (\tev)}\\
\hline
\multirow{2}{2.2cm}{Experiment/ energy} & \multirow{2}{*}{$Q~(e)$} & \phantom{999} & \multicolumn{2}{c}{DY $\gamma$ exchange} & \phantom{99} &\multicolumn{2}{c}{$\gamma\gamma$ fusion} \\ 
\cline{4-5}\cline{7-8}
 & & & LO & RES & &  LO & RES\\
\hline
\multirow{2}{2.2cm}{MoEDAL~\cite{MoEDAL:2023ost} $\sqrt{s}=13~\tev$} & 10 && 0.08 & \phantom{1}0.08\footnotemark[1] && 0.30 & \phantom{1}0.30\footnotemark[1] \\
& 15 && 0.22 & 0.25 && 0.74 & 0.88 \\
& 20   && 0.40 & 0.37 && 1.14 & 1.32 \\
& 25   && 0.58 & 0.63 && 1.50 & 1.73 \\
& 50   && 1.30 & 1.42 && 2.72 & 2.90 \\   
& 75   && 1.39 & 1.48 && 3.02 & 3.18 \\
& 100  && 1.42 & 1.51 && 3.17 & 3.31 \\   
& 125  && 1.43 & 1.53 && 3.26 & 3.40 \\
& 150  && 1.43 & 1.52 && 3.31 & 3.46 \\   
& 175  && 1.41 & 1.51 && 3.32 & 3.45 \\
& 200  && 1.39 & 1.50 && 3.31 & 3.44 \\
& 225  && 1.37 & 1.47 && 3.24 & 3.33 \\
& 250  && 1.34 & 1.44 && 3.16 & 3.25 \\
& 275  && 1.29 & 1.38 && 3.00 & \phantom{1}3.00\footnotemark[2] \\
& 300  && 1.21 & 1.30 && 2.50 & \phantom{1}2.50\footnotemark[2] \\
& 325  && 1.07 & 1.12 && 2.00 & \phantom{1}2.00\footnotemark[2] \\
& 350  && 0.90 & 0.93 && 1.50 & \phantom{1}1.50\footnotemark[2] \\
& 375  && 0.50 & \phantom{1}0.50\footnotemark[2] && 1.50 & \phantom{1}1.50\footnotemark[2] \\
\hline
\multirow{2}{2.2cm}{ATLAS~\cite{ATLAS:2023esy} $\sqrt{s}=13~\tev$} & 20 && 1.4 & 1.5 && 2.1 & 2.4 \\
& 40   && 1.8 & 1.9 && 2.8 & 3.0 \\
& 60   && 1.9 & 2.0 && 2.9 & 3.1 \\
& 80   && 1.8 & 1.9 && 2.8 & 3.0 \\
& 100  && 1.7 & 1.8 && 2.5 & \phantom{1}2.5\footnotemark[2] \\

\hline
\multirow{2}{2.2cm}{MoEDAL~\cite{MoEDAL:2021mpi} $\sqrt{s}=8~\tev$} & 15 && 0.07 & 0.09 && -- & -- \\
& 20   && 0.12  & 0.16 && -- & -- \\
& 25   && 0.19 & 0.27 && -- & --\\
& 50   && 0.56 & 0.62 && -- & -- \\   
& 75   && 0.58 & 0.64 && -- & -- \\
& 100  && 0.55  & 0.62 && -- & -- \\     
& 125  && 0.50 & 0.56 && -- & -- \\
& 130  && 0.49 & 0.56 && -- & -- \\
& 140  && 0.47 & 0.53 && -- & -- \\
& 145  && 0.47 & 0.53 && -- & --  \\
& 150  && 0.46 & 0.52 && -- & -- \\   
& 175  && 0.40 & 0.46 && -- & -- \\
\hline
\multirow{2}{2.2cm}{ATLAS~\cite{ATLAS:2015tyu} $\sqrt{s}=8~\tev$} 
& 10 && 0.49 & \phantom{1}0.49\footnotemark[1]  && --  & -- \\
& 20 && 0.78 & 0.81 && --  & -- \\
& 40 && 0.92 & 0.95 && --  & -- \\
& 60 && 0.88 & 0.92 && --  & -- \\

\hline\hline
\end{tabular}
\footnotetext[1]{Resummation is valid for $Q\gtrsim11e$, so there is no change in the mass limit for $Q=10e$.}
\footnotetext[2]{There is no experimental sensitivity for HECO masses higher than this value.}
\end{table*}

\begin{table*}[ht]
\caption{95\% CL experimental mass limits for spin-\hf HECOs with (RES) and without (LO) resummation techniques to calculate production cross sections. Reported bounds are not necessarily extracted from the same dataset size.}
\label{tab:limits-fermion}
\centering
\begin{tabular}{ c r c c c c c c c c c }
\hline\hline
\multicolumn{11}{ c }{Experimental lower limits at 95\% CL on spin-\hf HECO mass (\tev)}\\
\hline
\multirow{2}{2.2cm}{Experiment/ energy} & \multirow{2}{*}{$Q~(e)$} & \phantom{999} & \multicolumn{2}{c}{DY $\gamma$ exchange} & \phantom{99} & \multicolumn{2}{c}{DY $\gamma$/$Z^0$ exchange} & \phantom{99} &\multicolumn{2}{c}{$\gamma\gamma$ fusion} \\ 
\cline{4-5}\cline{7-8}\cline{10-11}
 & & & LO & RES & & LO & RES & & LO & RES\\
\hline
\multirow{2}{2.2cm}{MoEDAL~\cite{MoEDAL:2023ost} $\sqrt{s}=13~\tev$} 
& 10  && 0.29 & \phantom{1}0.29\footnotemark[1] && 0.32 & \phantom{1}0.32\footnotemark[1] && 0.41 & \phantom{1}0.41\footnotemark[1] \\
& 15  && 0.62 & 0.72 && 0.62 & 0.80 && 0.95 & 1.21 \\
& 20  && 0.92 & 1.08 && 0.93 & 1.15 && 1.38 & 1.68 \\
& 25  && 1.19 & 1.36 && 1.17 & 1.45 && 1.80 & 2.13 \\
& 50  && 1.85 & 2.02 && 1.84 & 2.10 && 3.00 & 3.27 \\   
& 75  && 1.94 & 2.10 && 1.93 & 2.16 && 3.25 & 3.51 \\
& 100 && 1.98 & 2.14 && 1.97 & 2.08 && 3.39 & 3.66 \\   
& 125 && 2.00 & 2.15 && 1.99 & 2.10 && 3.48 & 3.72 \\
& 150 && 2.00 & 2.16 && 1.98 & 2.10 && 3.52 & 3.75 \\   
& 175 && 1.98 & 2.13 && 1.97 & 2.17 && 3.53 & 3.73 \\
& 200 && 1.94 & 2.10 && 1.94 & 2.13 && 3.50 & 3.69 \\
& 225 && 1.90 & 2.03 && 1.90 & 2.08 && 3.44 & 3.60 \\
& 250 && 1.83 & 1.95 && 1.84 & 2.02 && 3.00 & \phantom{1}3.00\footnotemark[2]\\
& 275 && 1.74 & 1.84 && 1.74 & 1.88 && 3.00 & \phantom{1}3.00\footnotemark[2]\\
& 300 && 1.61 & 1.70 && 1.62 & 1.75 && 2.50 & \phantom{1}2.50\footnotemark[2]\\
& 325 && 1.38 & 1.46 && -- & -- && 2.00 & \phantom{1}2.00\footnotemark[2]\\
& 350 && 1.00 & \phantom{1}1.00\footnotemark[2] && -- & -- && 1.50 & \phantom{1}1.50\footnotemark[2] \\
& 375 && 0.20 & \phantom{1}0.20\footnotemark[2] && -- & -- && 1.50 & \phantom{1}1.50\footnotemark[2] \\
& 400 && -- & -- && -- & -- && 1.50 & \phantom{1}1.50\footnotemark[2]\footnotemark[3] \\
\hline
\multirow{2}{2.2cm}{ATLAS~\cite{ATLAS:2019wkg,ATLAS:2023esy} $\sqrt{s}=13~\tev$} & 20 && 1.83 & 2.02 && 1.8 & 1.9 &  & 2.5 &  2.7  \\
& 40   && 2.05 & 2.22 && 2.2 & 2.3 && 3.1 & 3.4 \\
& 60   && 2.00 & 2.18 && 2.2 & 2.4 && 3.1 & 3.4 \\
& 80   && 1.86 & 2.02 && 2.1 & 2.2 && 3.0 & \phantom{1}3.0\footnotemark[2] \\
& 100  && 1.65 & 1.80 && 1.9 & 2.1 && 2.5 & \phantom{1}2.5\footnotemark[2] \\
\hline
\multirow{2}{2.2cm}{MoEDAL~\cite{MoEDAL:2021mpi} $\sqrt{s}=8~\tev$} & 15 && 0.18 & 0.24 && 0.17 & 0.24 && -- & -- \\
& 20   && 0.28 & 0.36 && 0.31 & 0.36 && -- & -- \\
& 25   && 0.44 & 0.55 && 0.44 & 0.53 && -- & -- \\
& 50   && 0.78 & 0.88 && 0.78 & 0.87 && -- & -- \\
& 75   && 0.78 & 0.88 && 0.78 & 0.84 && -- & -- \\
& 100  && 0.73 & 0.84 && 0.71 & 0.80 && -- & -- \\
& 125  && 0.66 & 0.75 && 0.64 & 0.72 && -- & -- \\
& 130  && 0.64 & 0.74 && 0.62 & 0.70 && -- & -- \\
& 140  && 0.58 & 0.68 && 0.62 & 0.69 && -- & -- \\
& 145  && 0.52 & 0.66 && 0.51 & 0.60 && -- & -- \\
& 150  && 0.50 & 0.63 && 0.58 & 0.66 && -- & -- \\
\hline
\multirow{2}{2.2cm}{ATLAS~\cite{ATLAS:2015tyu} $\sqrt{s}=8~\tev$} & 10 && 0.78 & \phantom{1}0.78\footnotemark[1] && -- & -- && -- & --  \\
& 20 && 1.05 & 1.14 && -- & -- && -- & -- \\
& 40 && 1.16 & 1.25 && -- & -- && -- & -- \\
& 60 && 1.07 & 1.15 && -- & -- && -- & -- \\
\hline\hline
\end{tabular}
\footnotetext[1]{Resummation is valid for $Q\gtrsim11e$, so there is no change in the mass limit for $Q=10e$.}
\footnotetext[2]{There is no experimental sensitivity for HECO masses higher than this value.}
\footnotetext[3]{The excluded mass range is 0.5--1.5~\tev.}
\end{table*}

\clearpage

\begin{figure}[ht]
    \includegraphics[width=0.6\linewidth]{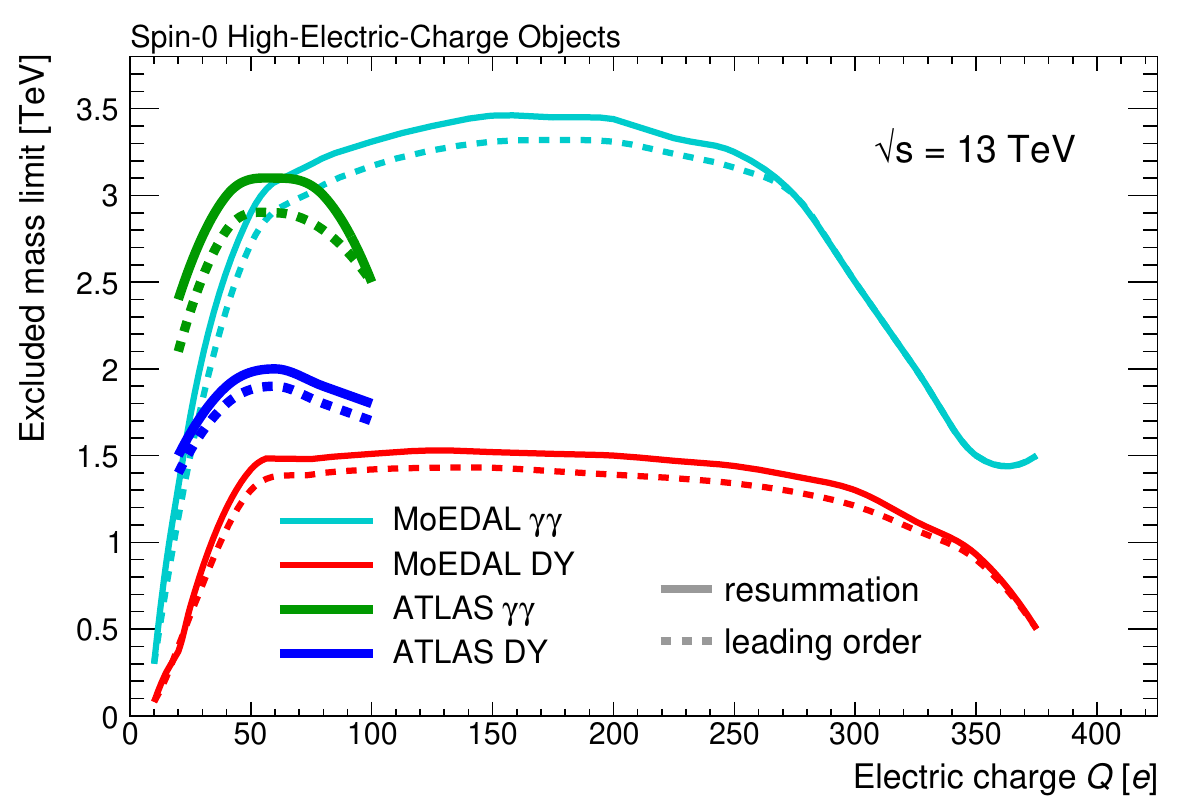}
    \caption{Excluded lower mass limits for spin-0 HECOs from ATLAS~\cite{ATLAS:2023esy} and MoEDAL~\cite{MoEDAL:2023ost} at $\sqrt{s}=13~\tev$ at leading order (dashed lines) and re-calculated with resummation (continuous lines). The dip around 350~\gev is an artifact of the curve smoothing.}
    \label{fig:hecos-resum-spinzero}
\end{figure}

\begin{figure}[ht]
    \includegraphics[width=0.6\linewidth]{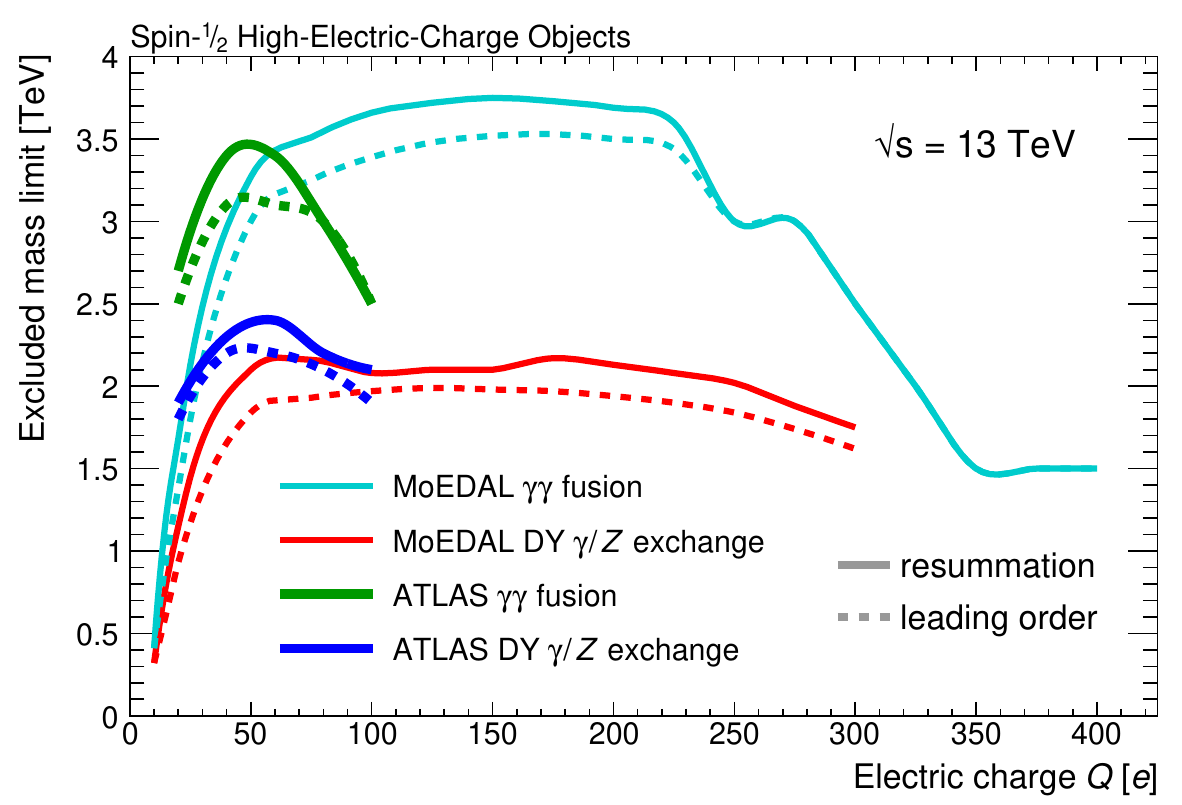}
    \caption{Excluded lower mass limits for spin-\hf HECOs from ATLAS~\cite{ATLAS:2023esy} and MoEDAL~\cite{MoEDAL:2023ost} at $\sqrt{s}=13~\tev$ at leading order (dashed lines) and re-calculated with resummation (continuous lines). The dips around 250~\gev and 350~\gev are artifacts of the curve smoothing.}
    \label{fig:hecos-resum-spinhalf}
\end{figure}

An overview of the spin-0 and spin-\hf HECO mass limits versus their electric charge is given in Fig.~\ref{fig:hecos-resum-spinzero} and Fig.~\ref{fig:hecos-resum-spinhalf}, respectively. 
The bounds acquired with resummation are compared to those obtained by the experimental collaborations at tree level at collision energy of $\sqrt{s}=13~\tev$. The enhancement of the mass limits is evident, except for high charges, where the MoEDAL sensitivity is limited for high masses, therefore the increased predicted cross section does not affect the mass limits.

\newpage

Before closing the section, we stress once again that the above results were formally derived in the Feynman gauge $\lambda=1$, which is a ``preferred'' (physical) gauge according to the PT framework. Nonetheless, for the sake of argument, one might consider the effects of working in a different gauge, e.g.\ the Landau one $\lambda \to \infty$, on the physical results. In this respect, it should be noted that the cross sections depend on the wave-function renormalization $\mathcal Z^\star$ at the UV fixed point, \eqref{rangeom}, which, unlike the photon-wavefunction renormalization $\omega^\star$ and the renormalized scalar self-interaction coupling $H^\star$, depends on the gauge parameter to first order in the small quantity $g^2/h$,  This would result in increased values of the DY and PF production cross sections by 20\% and 44\%, respectively, compared to the Feynman gauge. Given that mass bounds are extracted from the logarithms of the cross sections, the above 
effect would have negligible consequences for the scalar HECO bounds.  

\section{Conclusions and outlook}\label{sec:6}

We have extended the resummation analysis of \cite{Alexandre:2023qjo}, based on the Dyson-Schwinger approach for spin-\hf HECOs, to the case of scalar (spin-$0$) HECOs. The method relies on the existence of non-trivial UV stable fixed points, at which an effective Lagrangian of their (strong) electromagnetic interactions is constructed, which is then implemented in the appropriate UFO models, used in the \MAD event generator, in order to extract production cross sections, and through them, reliable mass bounds by comparison with the experimental constraints.  

Although in the fermion-HECO case, the existence of such non-trivial UV fixed points emerged solely from the interaction of the HECO with the photon field, in the scalar-HECO case, the HECO-photon interaction alone is not sufficient to guarantee the existence of a UV non-trivial fixed point. One needs a minimum self-interaction of the scalar HECO fields among themselves \eqref{gvsh}, in order to obtain a finite non-trivial UV fixed point. On saturating the lower bound \eqref{gvsh}, i.e.\ assuming \eqref{satgvsh},
we have considered the effective field theory of scalar HECOs evaluated at that fixed point, and compared its predictions with the results of the experimental searches of ATLAS~\cite{ATLAS:2015tyu,ATLAS:2019wkg,ATLAS:2023esy} and MoEDAL experiments~\cite{MoEDAL:2021mpi,MoEDAL:2023ost}. Following the assumptions in~\cite{Song:2021vpo}, we considered only the coupling of the scalar HECOs to photons in the DY production process. 
The model we considered, therefore, for scalar HECOs was a strongly coupled scalar electrodynamics, but with (charged) scalar self interactions.
As in the fermionic-HECO case, we also found that the scalar-HECO resummed theory leads to higher cross sections, as compared to the tree-level estimated ones, used in the searches, which in turn implies reliable scalar HECO mass lower limits,, which are larger (by up to about 30 \%)  as compared to those extracted from tree-level DY and PF processes.

Further work to be done in this respect pertains to considering a DS improved resummation, obtained after making more detailed assumptions for the dressed quantities, e.g. including momentum-dependent quantum corrections, which is equivalent to include more complete sets of Feynman graphs in the resummation.
Hopefully the results will not be quantitatively very  different from the current case, which would be a rather nice consistency check of the convergence of the resummation process, as far as higher-loop improvements are concerned. We hope to be able to address such issues in a future publication.

\section*{Acknowledgements}

The research of J.A.\ and N.E.M.\ is supported in part by the UK Science and Technology Facilities research Council (STFC) under the research grants ST/T000759/1 and ST/X000753/1, while that of V.A.M.\ and E.M.\ is supported by the Generalitat Valenciana via the Excellence Grant Prometeo CIPROM/2021/073, by the Spanish MCIU / AEI / 10.13039/501100011033 and the European Union / FEDER via the grant PID2021-122134NB-C21, and by the Spanish MCIU/AEI via the Severo Ochoa project CEX2023-001292-S.

\appendix

\section{Loop integrals}\label{app:loop}

We use dimensional regularisation in dimension $d=4-\epsilon$, $\epsilon \to 0^+$, where $g\to g k^{\epsilon/2}$, $h\to h k^{\epsilon/2}$.
We have then
\be
\int\frac{d^dp}{(2\pi)^d}\frac{1}{p^2-A^2}=-\frac{i\pi^{d/2}}{(2\pi)^d}A^{2-\epsilon}\Gamma(-1+\epsilon/2)
=\frac{iA^{2-\epsilon}}{8\pi^2\epsilon}~+~A^{2-\epsilon}f_1~,
\ee
where $f_1$ is dimensionless and finite, and $\Gamma$ is the gamma function. As a consequence we have
\be
\int\frac{d^dp}{(2\pi)^d}\frac{1}{(p^2-A^2)^2}=\frac{d}{d(A^2)}\int\frac{d^dp}{(2\pi)^d}\frac{1}{p^2-A^2}
=\frac{iA^{-\epsilon}}{8\pi^2\epsilon}~+~A^{-\epsilon}f_2~,
\ee
where $f_2$ is dimensionless and finite. Also,
\be
\int\frac{d^dp}{(2\pi)^d}\frac{1}{p^2}=\lim_{A\to0}\int\frac{d^dp}{(2\pi)^d}\frac{1}{p^2-A^2}=0~,
\ee
and
\be
\int\frac{d^dp}{(2\pi)^d}\frac{1}{p^2(p^2-A^2)}
=A^{-2}\int\frac{d^dp}{(2\pi)^d}\left(\frac{1}{p^2}-\frac{1}{p^2-A^2}\right)=-\frac{iA^{-\epsilon}}{8\pi^2\epsilon}~+~A^{-\epsilon}f_1~.
\ee
An infrared divergence is regularised in $d'=4+\epsilon$, with $\epsilon>0$, such that
\be
\int\frac{d^{d'}p}{(2\pi)^{d'}}\frac{1}{p^4}=\lim_{A\to0}\int\frac{d^{d'}p}{(2\pi)^{d'}}\frac{1}{(p^2-A^2)^2}
=\lim_{A\to0}\left(-\frac{iA^{+\epsilon}}{8\pi^2\epsilon}+A^{+\epsilon}f_3\right)=0~,
\ee
where $f_3$ is dimensionless and finite.

\begin{figure}[ht]
    \includegraphics[width=0.8\linewidth]{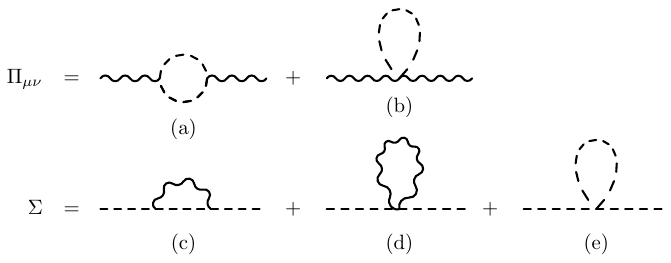}
    \caption{The photon (a+b) and scalar (c+d+e) one-loop self energies. In the resummation we propose, all the vertices and 
    internal lines are replaced by the dressed ones.}
    \label{PiSigma}
\end{figure}

\begin{figure}[ht]
    \includegraphics[width=0.8\linewidth]{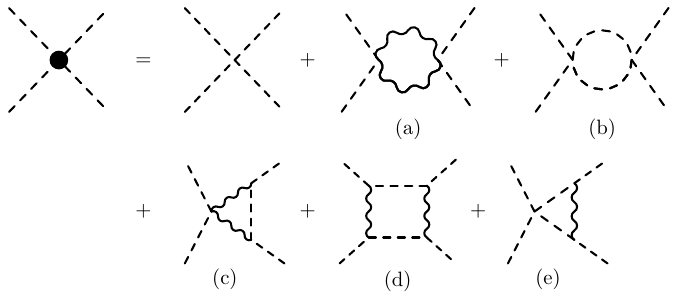}
    \caption{One-loop corrections to the scalar self-coupling. Each graph contributes three times, 
    due to the three different channels for the external momenta. In the resummation we propose, all the vertices and 
    internal lines are replaced by the dressed ones. Each graph is discussed in the text.}
    \label{H}
\end{figure}

\section{Polarization tensor of the gauge bosons}\label{sec:poltens}

The polarisation tensor is defined as 
\be
i\Pi_{\mu\nu}(q)\equiv D^{-1}_{\mu\nu}(q)-D^{-1}_{(bare)\mu\nu}(q)~,
\ee
which is transverse, and is denoted
\be
\Pi_{\mu\nu}(q)\equiv -\omega\left(q^2\eta_{\mu\nu}-q_\mu q_\nu\right)~.
\ee
The one-loop polarization tensor does not get contributions from the scalar self interactions, but only from the gauge coupling.
It results from the sum of graph (a) and graph (b) on Fig.~\ref{PiSigma}) and reads \cite{Schwartz:2014sze}
\be
\omega^{(1)}=\frac{g^2}{24\pi^2}\frac{1}{\epsilon}\left(\frac{k}{m}\right)^\epsilon+\mbox{finite}~.
\ee
The resummation considered in this article consists in replacing the bare quantities in the loop integral by the dressed ones,
which results in 
\be
\omega=\frac{g^2}{24\pi^2{\mathcal Z}^2}\frac{1}{\epsilon}\left(\frac{k{\mathcal Z}}{M}\right)^\epsilon+\mbox{finite}~.
\ee

\section{Scalar self energy}\label{scalarself}

The scalar self energy is given by
\be
i\Sigma(p)\equiv G^{-1}(p)-G_{(bare)}^{-1}(p)=({\mathcal Z}^2-1)p^2-(M^2-m^2)~,
\ee
and has contributions from three graphs, shown in Fig.~\ref{PiSigma}. 
The graphs (d) and (e) do not depend on the external momentum, such that they do not contribute to the correction ${\mathcal Z}$. 
The graph (d) does not contribute since it leads to a quadratic divergence, not seen by dimensional regularisation
\bea
\Sigma_{(d)}&=&-2g^2\eta_{\mu\nu}{\mathcal Z}^2k^\epsilon
\int\frac{d^dq}{(2\pi)^d}\frac{\eta^{\mu\nu}+\xi q^\mu q^\nu/q^2}{(1+\omega)q^2}\\
&=&-2g^2{\mathcal Z}^2k^\epsilon\frac{4+\xi}{1+\omega}\int\frac{d^dq}{(2\pi)^d}\frac{1}{q^2}=0\nonumber~,
\eea
where $\xi\equiv(1+\omega-\lambda)/\lambda$.
The contribution from the graph (e) is
\bea
\Sigma_{(e)}(0)&=&-\frac{Hk^\epsilon}{{\mathcal Z}^2}\int\frac{d^dq}{(2\pi)^d}\frac{1}{q^2-(M/{\mathcal Z})^2}\\
&=&\frac{H}{{\mathcal Z}^2}\frac{i}{8\pi^2\epsilon}\left(\frac{M}{{\mathcal Z}}\right)^2\left(\frac{k{\mathcal Z}}{M}\right)^\epsilon
+\mbox{finite}~.\nonumber
\eea 
The contribution (c) to the scalar self energy is
\be
\Sigma_{(c)}(p)=-\frac{{\mathcal Z}^2g^2k^\epsilon}{1+\omega}\int\frac{d^dq}{(2\pi)^d}\left(\eta_{\mu\nu}+\xi\frac{q_\mu q_\nu}{q^2}\right)
\frac{(q+2p)^\mu(q+2p)^\nu}{q^2[{\mathcal Z}^2(p+q)^2-M^2]}~,
\ee
such that the mass correction is given by
\bea
\Sigma_{(c)}(0)+\Sigma_{(e)}(0)
&=&\frac{H}{{\mathcal Z}^2}\frac{i}{8\pi^2\epsilon}\left(\frac{M}{{\mathcal Z}}\right)^2\left(\frac{k{\mathcal Z}}{M}\right)^\epsilon
-ig^2\frac{1+\xi}{1+\omega}k^\epsilon\int\frac{d^dq}{(2\pi)^d}\frac{1}{q^2-M^2/{\mathcal Z}^2}\\
&=&\frac{H}{{\mathcal Z}^2}\frac{i}{8\pi^2\epsilon}\left(\frac{M}{{\mathcal Z}}\right)^2\left(\frac{k{\mathcal Z}}{M}\right)^\epsilon
+\frac{ig^2}{8\pi^2\lambda}\left(\frac{k{\mathcal Z}}{M}\right)^\epsilon\left(\frac{M}{{\mathcal Z}}\right)^2+\mbox{finite}~,\nonumber
\eea
which leads to
\be
M^2-m^2=\frac{1}{8\pi^2\epsilon}\left(\frac{g^2}{\lambda}+\frac{H}{{\mathcal Z}^2}\right)\left(\frac{M}{{\mathcal Z}}\right)^2
\left(\frac{k{\mathcal Z}}{M}\right)^\epsilon+\mbox{finite}~.
\ee
The derivative correction is given by
\bea
8({\mathcal Z}^2-1)=i\left.\frac{\partial^2\Sigma_3}{\partial p^\rho\partial p_\rho}\right|_{p=0}
&=&-\frac{8g^2k^\epsilon}{1+\omega}\int\frac{d^dq}{(2\pi)^d}\left(\frac{(1+\xi)(M/{\mathcal Z})^2}{[q^2-(M/{\mathcal Z})^2]^3}
-\frac{2(1+\xi)}{[q^2-(M/{\mathcal Z})^2]^2}+\frac{4+\xi}{q^2[q^2-(M/{\mathcal Z})^2]}\right)\\
&=&\frac{8g^2k^\epsilon}{1+\omega}\int\frac{d^dq}{(2\pi)^d}\left(\frac{2(1+\xi)}{[q^2-(M/{\mathcal Z})^2]^2}
-\frac{(4+\xi)({\mathcal Z}/M)^2}{q^2-(M/{\mathcal Z})^2}\right)+\mbox{finite}\nonumber\\
&=&\frac{g^2}{\pi^2\epsilon}\left(\frac{k{\mathcal Z}}{M}\right)^\epsilon
\left(\frac{3}{1+\omega}-\frac{1}{\lambda}\right)+\mbox{finite}~.\nonumber
\eea

\section{Scalar self interaction}\label{app:selfinter}

The diagrams contributing to the scalar self interaction are shown in Fig.~\ref{H}, 
and the total contribution to the scalar self coupling is denoted by
\be
H=h+H_{(a)}+H_{(b)}+H_{(c)}+H_{(d)}+H_{(e)}~,
\ee
where we will ignore finite terms. All these graphs count three times, due to the different number of channels for the incoming/outgoing momenta.

The graph (a) has an infrared divergence but it does not contribute, since
\bea
H_{(a)}&\propto&\eta^{\mu\nu}\eta^{\rho\sigma}\int\frac{d^{d'}q}{(2\pi)^{d'}}
\frac{1}{[(1+\omega)q^2]^2}\left(\eta_{\mu\rho}+\xi\frac{q_\mu q_\rho}{q^2}\right)\left(\eta_{\nu\sigma}+\xi\frac{q_\nu q_\sigma}{q^2}\right)\\
&=&\frac{4+2\xi+\xi^2}{(1+\omega)^2}\int\frac{d^{d'}q}{(2\pi)^{d'}}\frac{1}{q^4}=0\nonumber~.
\eea
From a physical point of view, the cancellation of the infrared divergence is explained by soft photons emitted/absorbed by the external scalar legs.
The improved one-loop diagram from the graph (b) is given by
\be
-ik^\epsilon H_{(b)}=3(-ik^\epsilon H)^2\int\frac{d^dq}{(2\pi)^d}\frac{i^2}{({\mathcal Z}^2p^2-M^2)^2}~,
\ee
where no symmetry factor is present since the scalar is charged \cite{Itzykson:1980rh}. We have then
\be
H_{(b)}=-\frac{3H^2}{{\mathcal Z}^4}\frac{1}{8\pi^2\epsilon}\left(\frac{k{\mathcal Z}}{M}\right)^\epsilon~.
\ee
The graph (c) is given by
\bea
-ik^\epsilon H_{(c)}&=&3\times2ig^2{\mathcal Z}^2(ig{\mathcal Z})^2\eta^{\mu\nu}k^{2\epsilon}\int\frac{d^dq}{(2\pi)^d}
\frac{(-i)^2q^\rho q^\sigma}{[(1+\omega)q^2]^2}\left(\eta_{\mu\rho}+\xi\frac{q_\mu q_\rho}{q^2}\right)\left(\eta_{\nu\sigma}+\xi\frac{q_\nu q_\sigma}{q^2}\right)
\frac{i}{{\mathcal Z}^2q^2-M^2}\\
&=&-6g^4{\mathcal Z}^2\left(\frac{1+\xi}{1+\omega}\right)^2k^{2\epsilon}\int\frac{d^dq}{(2\pi)^d}\frac{1}{q^2(q^2-(M/{\mathcal Z})^2)}\nonumber\\
&=&-\frac{6g^4{\mathcal Z}^2}{\lambda^2}\frac{ik^\epsilon}{8\pi^2\epsilon}\left(\frac{k{\mathcal Z}}{M}\right)^\epsilon+\mbox{finite}~,\nonumber
\eea
such that 
\be
H_{(c)}=\frac{3g^4{\mathcal Z}^2}{4\pi^2\lambda^2\epsilon}\left(\frac{k{\mathcal Z}}{M}\right)^\epsilon~.
\ee
The graph (d) is given by 
\bea
-ik^\epsilon H_{(d)}&=&3(ig{\mathcal Z})^4k^{2\epsilon}\int\frac{d^dq}{(2\pi)^d}
\frac{(-i)^2q^\mu q^\nu q^\rho q^\sigma}{[(1+\omega)q^2]^2}\left(\eta_{\mu\nu}+\xi\frac{q_\mu q_\nu}{q^2}\right)
\left(\eta_{\rho\sigma}+\xi\frac{q_\rho q_\sigma}{q^2}\right)\frac{i^2}{({\mathcal Z}^2q^2-M^2)^2}\\
&=&3g^4\left(\frac{1+\xi}{1+\omega}\right)^2k^{2\epsilon}\int\frac{d^dq}{(2\pi)^d}\frac{1}{[q^2-(M/{\mathcal Z})^2]^2}\nonumber\\
&=&\frac{3g^4}{\lambda^2}\frac{ik^\epsilon}{8\pi^2\epsilon}\left(\frac{k{\mathcal Z}}{M}\right)^\epsilon+\mbox{finite}~,\nonumber
\eea
such that 
\be
H_{(d)}=-\frac{3g^4}{8\pi^2\lambda^2\epsilon}\left(\frac{k{\mathcal Z}}{M}\right)^\epsilon~.
\ee
Finally, the graph (e) is given by 
\bea
-ik^\epsilon H_{(e)}&=&-3iH(ig{\mathcal Z})^2k^{2\epsilon}\int\frac{d^dq}{(2\pi)^d}\frac{i^2q^\mu q^\nu}{({\mathcal Z}^2q^2-M^2)^2}
\frac{-i}{(1+\omega)q^2}\left(\eta_{\mu\nu}+\xi\frac{q_\mu q_\nu}{q^2}\right)\\
&=&-\frac{3Hg^2}{{\mathcal Z}^2}\frac{1+\xi}{1+\omega}k^{2\epsilon}\int\frac{d^dq}{(2\pi)^d}\frac{1}{[q^2-(M/{\mathcal Z})^2]^2}\nonumber\\
&=&-\frac{3Hg^2}{{\mathcal Z}^2\lambda}\frac{ik^\epsilon}{8\pi^2\epsilon}\left(\frac{k{\mathcal Z}}{M}\right)^\epsilon+\mbox{finite}~,\nonumber
\eea
such that 
\be
H_{(e)}=\frac{3Hg^2}{8\pi^2{\mathcal Z}^2\lambda\epsilon}\left(\frac{k{\mathcal Z}}{M}\right)^\epsilon~.
\ee

\section{UFO model validation}\label{validation_appendix}
To ensure the validity of the UFO model for HECOs described in Section~\ref{sec:3}, we compared cross-section values from analytical calculations using the \Math package \Feyncalc9.3.1~\cite{Feyncalc} with those calculated by \madamc. This validation step is crucial for verifying the consistency and accuracy of the UFO models, thereby enhancing confidence in their predictive performance.

Before validating the UFO model which incorporates resummation effects, we first confirmed the accuracy of tree-level estimates by examining the total cross section for HECO pair production. We compared the \MAD results with the corresponding analytical results from \Math. For this validation, we focused on direct collisions of initial up-quarks or photons without using PDFs, applying the \texttt{no-PDF} option in \MAD.

The results of the comparison between \MAD and \Math for DY and PF processes involving $u\bar{u}$ quarks and photons, respectively, producing $\h \hb $ at tree level are presented in Table~\ref{tab:val-lo}. The UFO/theory ratio is close to unity, confirming that the simulation accurately reflects the theoretical predictions at tree level.

\begin{table*}[ht]
\caption{Cross-section of spin-0 HECO production via the DY $u \Bar{u} \to \h \hb$ and the PF $\gamma \gamma \to \h \hb$ processes at $\sqrt{s}=13~\tev$ obtained with \MAD by importing the leading-order UFO model and with theoretical calculations. A HECO mass $M= 1~\tev$ is assumed. The simulation/theory ratio is also listed.}
\label{tab:val-lo}
\centering
\begin{tabular}{ c c c c c c c c c }
\hline\hline
\multicolumn{9}{ c }{Validation of UFO model at tree level @ $\sqrt{s}=13~\tev$, $M = 1~\tev$}\\
\hline
\multirow{2}{*}{$Q\,(e)$}&&\multicolumn{3}{c}{DY $u \Bar{u} \to \h\hb$}&&\multicolumn{3}{c}{PF $\gamma \gamma \to \h\hb$}\\ \cline{3-5}\cline{7-9}
 && $\sigma_\MAD$ (pb) & $\sigma_\Math$  (pb)&  UFO/Theory & &$\sigma_\MAD$ (pb) & $\sigma_\Math$  (pb)&  UFO/Theory\\
\hline
\phantom{0}20 & \phantom{99} & $8.418\times10^{-3}$ & $8.382\times10^{-3}$ & 1.004 & \phantom{99} & $1.263\times 10^2$  & $1.255\times 10^2$ & 1.006 \\
\phantom{0}60 && $7.588\times10^{-2}$ & $7.543\times10^{-2}$ & 1.006 && $1.023\times 10^4$ &  $1.017\times 10^4$  & 1.006 \\
100 && 0.2107 & 0.2095 & 1.005 && $7.897\times 10^4$ & $7.845 \times 10^4$& 1.005 \\
140 && 0.4132 & 0.4107 & 1.006 && $3.026\times 10^5$  & $3.013\times 10^5$ & 1.006 \\
180 && 0.6829 & 0.6789 & 1.005 && $8.279\times 10^5$& $8.235\times 10^5$  & 1.006 \\
220 && 1.019\phantom{0}  & 1.014\phantom{0}  & 1.006 && $1.846\times 10^6$ & $1.838\times 10^6$ & 1.006 \\
\hline\hline
\end{tabular}
\end{table*}

Then, we validate the actual UFO model that includes DS resummation effects, applying the same methodology used for tree-level validation. Table~\ref{tab:val-res} lists  cross-section values for the DY and PF processes with resummation effects implemented following the Feynman rules \eqref{feynrul}. In both processes, the results from the UFO model show good agreement with theoretical predictions. This ensures that these tools can be reliably used for simulating high-energy particle collisions with the implemented resummation effects.

\begin{table*}[ht]
\caption{Cross-section of spin-0 HECO production via the DY $u \Bar{u} \to \h \hb$ and the PF $\gamma \gamma \to \h \hb$ processes at $\sqrt{s}=13~\tev$ obtained with \MAD by importing the resummation UFO model and with theoretical calculations. Resummation with $\Lambda= 2~\tev$ is assumed. The simulation/theory ratio is also listed.}
\label{tab:val-res}
\centering
\begin{tabular}{ c c c c c c c c c }
\hline\hline
\multicolumn{9}{ c }{Validation of UFO model with resummation @ $\sqrt{s}=13~\tev$, $\Lambda= 2~\tev$}\\
\hline
\multirow{2}{*}{$Q\,(e)$}&&\multicolumn{3}{c}{DY $u \Bar{u} \to \h\hb$}&&\multicolumn{3}{c}{PF $\gamma \gamma \to \h\hb$}\\ \cline{3-5}\cline{7-9}
 && $\sigma_\MAD$ (pb) & $\sigma_\Math$  (pb)&  UFO/Theory & &$\sigma_\MAD$ (pb) & $\sigma_\Math$  (pb)&  UFO/Theory\\
\hline
\phantom{0}20 & \phantom{99} & $0.01396$ & $0.01388$ & 1.006 & \phantom{99} & $3.458\times10^{2}$  & $3.439\times10^{2}$  & 1.006 \\
\phantom{0}60 && $0.1149\phantom{0}$  & $0.1144\phantom{0}$ &   1.004 && $2.320\times10^{4}$ & $2.309\times10^{4}$ & 1.005 \\
100 && $0.3149\phantom{0}$  & $0.3131\phantom{0}$ &  1.006 && $1.750\times10^{5}$  & $1.741\times10^{4}$&   1.005 \\
140 && $0.6151\phantom{0}$ & $0.6113\phantom{0}$ & 1.006 && $6.677\times10^{5}$ & $6.643\times10^{5}$& 1.005 \\
180 && 1.015\phantom{00} & 1.009\phantom{00}   & 1.006 && $1.820\times10^{6}$ & $1.810\times10^{5}$    & 1.006 \\
220 && 1.514\phantom{00} & 1.506\phantom{00} & 1.005 && $4.058\times10^{6}$ & $4.034\times10^{6}$ & 1.006 \\
\hline\hline
\end{tabular}
\end{table*}

\bibliographystyle{apsrev4-1}
\bibliography{scalarHECO}

\end{document}